\documentclass[a4paper,11pt]{article}

\usepackage{jheppub} 

\usepackage{fancyhdr}
\usepackage{graphicx}
\usepackage{epstopdf}
\usepackage{subfigure}

\usepackage{axodraw4j}
\usepackage{color}

\usepackage{braket}
\usepackage{slashed}

\newcommand{\be}{\begin{equation}}
\newcommand{\ee}{\end{equation}}
\newcommand{\bea}{\begin{eqnarray}}

\newcommand{\eea}{\end{eqnarray}}

\newcommand{\refe}[1]{Eqn.~(\ref{#1})}

\newcommand{\Rmnum}[1]{\expandafter\@slowromancap\romannumeral #1@}

\begin{document} 

 \title{\boldmath Confronting Higgs couplings from D-term extensions\\ and Natural SUSY at the LHC and ILC}

 \begin{flushright}
 WITS-CTP-137\\
DESY 14-090
 \end{flushright}

\author[\bigstar,\spadesuit]{ Moritz McGarrie,}
\author[\clubsuit,\diamondsuit]{Gudrid Moortgat-Pick}
\author[\clubsuit]{and Stefano Porto}


\affiliation[\bigstar]{School of Physics and Centre for Theoretical Physics,
University of the Witwatersrand, \\Johannesburg, WITS 2050, South Africa}

\affiliation[\spadesuit]{Institute of Theoretical Physics, Faculty of Physics, University of Warsaw,\\
ul. Ho\.za 69, 00-681 Warszawa, Poland}

\affiliation[\clubsuit]{II. Institut f\"{u}r Theoretische Physik, Universit\"{a}t Hamburg,\\
Luruper Chaussee 149, 22761 Hamburg, Germany}

\affiliation[\diamondsuit]{DESY, Deutsches Elektronen-Synchrotron,\\ Notkestra\ss{}e 85, D-22607 Hamburg, Germany}

\abstract{
Non-decoupling D-term extensions of the MSSM enhance
the tree-level Higgs mass compared to the MSSM, therefore relax fine-tuning and may allow lighter stops with rather low masses  
even without maximal mixing.
We present the anatomy of various non-decoupling D-term
extensions of the MSSM and explore the potential of the LHC and of the
International Linear Collider (ILC) to determine their
deviations in the Higgs
couplings with respect to the Standard Model. Depending on the mass of the heavier Higgs $m_H$,
such deviations may be constrained at the LHC and determined at the ILC. We evaluate the Higgs couplings in different models and study the prospects for a model distinction at the different stages of the ILC at $\sqrt{s}=$250, 500, 1000 GeV, including the full luminosity upgrade and compare it with the prospects at HL-LHC.
}

\maketitle
\flushbottom

\section{Introduction} \label{sec:intro}

The mass of the recently discovered scalar particle $m_h \sim 125.5$ GeV at the Large Hadron
Collider (LHC) \cite{Aad:2012tfa,Chatrchyan:2012ufa}, as well as its measured signal
strengths, within the current achievable precision, 
is consistent with the Higgs boson of the Standard Model (SM). 
In the context of Supersymmetry (SUSY), 
the observed Higgs
mass can be obtained within the minimal supersymmetric standard model
(MSSM), as well as a number of well-defined extensions of the MSSM based
on the two Higgs doublet model \cite{Gunion:2002zf}.  However, having
not yet observed supersymmetric particles at the LHC so far may provide
circumstantial evidence that the MSSM is 
fine-tuned to some degree, as the generation of such a heavy mass for the lightest CP-even Higgs often requires
heavy stops, posing a naturalness problem, or high stop mixings. 
  
Motivated by the aesthetic of naturalness and in the
endeavour to uniquely determine the Higgs sector and its scalar
potential,  in
this paper we explore a number of concrete and
well motivated extensions of the MSSM and
study to what degree they lead to deviations from the SM that are measurable 
at the LHC or at a future Higgs factory such as the International Linear Collider (ILC).

There are two main categories of extensions of the MSSM that may offer
extra contributions to the Higgs mass at tree level, thereby
improving fine-tuning. The first category is
given by $F$-term extensions of the MSSM, in which additional fields
interacting with the MSSM Higgs doublets - either gauge singlets as in
the NMSSM \cite{Fayet:1974pd,Nilles:1982dy,Frere:1983ag,Derendinger:1983bz,Durand:1988rg,Ellis:1988er} (for a review see \cite{Maniatis:2009re,Ellwanger:2009dp,Drees:2004jm})
or triplets \cite{Espinosa:1991wt,Espinosa:1992hp,Espinosa:1998re,Delgado:2013zfa} - 
raise the tree-level Higgs mass
via terms in the superpotential resulting in enhanced quartic couplings of the Higgs boson. 
The second category, that will be studied in this
work, is given by quiver or D-term extensions of the MSSM
\cite{Batra:2003nj,Maloney:2004rc}.  In these models, an MSSM gauge
group extension provides additional non-decoupling D-terms from the
K\"ahler potential, enhancing the tree level Higgs mass through extra
contributions to the Higgs quartic couplings. In particular, at a scale
above the TeV-scale, the extended gauge group under which the Higgs
fields are charged is broken to $SU(2)_{L}\times U(1)_Y$; the
additional D-terms originate from integrating out the heavier scalar
fields that participate in the breaking of the gauge groups. This
category of MSSM extension is appealing for a series of reasons
\cite{Batra:2003nj,Maloney:2004rc}: the electroweak scale remains stable
after running from higher energies, as there are no log-enhanced 1-loop
corrections to Higgs soft masses; additional contributions to
electroweak precision observables can be suppressed and gauge coupling
unification is not obviously spoiled.   
In addition, these models are consistent and compatible
with all frameworks of supersymmetry breaking, the Higgs enhancement
being largely independent of how
SUSY breaking effects are parameterised.\footnote{At low energies the model is often well described
by the MSSM plus an effective action. Therefore the soft terms can be
parameterised largely independent of the D-terms enhancement, if the
scale of diagonal gauge symmetry breaking
 is small enough.}

We consider gauge extended MSSM models in which the gauge group
features two copies of the electroweak gauge group $SU(2)\times U(1)$, $G_A$ in
site $A$ and $G_B$
in site $B$. At lower energies, at  a scale $\gtrsim 1$ TeV, $G_A\times G_B$ diagonally breaks to the SM electroweak group
$SU(2)_{L}\times U(1)_Y$. In this case, two
main classes of models can be identified. In the first class, which we
will refer to as the ``vector Higgs" case, the two Higgs doublets $H_u$
and $H_d$ are both charged either under $G_A$ or under $G_B$, transforming as a vector representation $(H_u,H_d)$ of $G_A\times G_B$. The second
class, the ``chiral Higgs" case, instead, has $H_u$ and $H_d$
charged under different copies of $SU(2)\times U(1)$ \cite{Randalltalk,Craig:2012bs}.

We supply an anatomy of these types of models, explore
whether they lead to
predictions that are experimentally testable at the LHC and the ILC, and use them as a predictive
guide concerning the stops masses and the trilinear $A_t$.

The approach we take here will be bottom-up in which we neglect effects from the renormalization group equations (RGE)
and focus on these extensions as deformations of the MSSM.  This
approach is complementary to that of \cite{Bharucha:2013ela}, for
instance, where a fully UV-complete two-loop spectrum generator is
used (and made publicly available, \cite{Staub:2008uz}) to analyse the
sparticle spectrum and Higgs physics of such a quiver model. Other
descriptions of quiver models as UV completions may be found, for
example in
\cite{Csaki:2001em,Cheng:2001an,Batra:2004vc,Delgado:2004pr,Medina:2009ey,Huo:2012tw,Randalltalk,DeSimone:2008gm,McGarrie:2010qr,Auzzi:2012dv,d2013fitting}.

The outline of this paper is as follows: in
section \ref{section:catalogue} we compare the minimisation conditions
and naturalness between the MSSM and some of its two-site quiver
extensions. 
  In section
\ref{section:HiggsatILC} we explore 
the LHC's and ILC's capabilities to resolve
such D-terms enhancements of the MSSM.  In section
\ref{section:conclusion} we discuss and conclude. In appendix \ref{app:generalDterms}
we supply a more general derivation of the D-terms for
Higgs bosons, squarks
and sleptons applicable to both chiral and vector Higgs models. In the appendices
\ref{app:HiggsSfermions} and \ref{app:HiggsSfermionschiral} we list
the mixing matrices of the Higgs sector for both vector-like and chiral
D-terms cases.  In appendix \ref{appendix:sfermionmatrix} we give the corrections
to the sfermion mass matrices of the MSSM, due to vector-like D-term
contributions. Appendix \ref{sec:unification} explores unification in these models as a guide to constrain the maximum possible size of D-terms.  Appendix \ref{sec:otherquivers} outlines  a holographic
3-site quiver extension that may also lead to non-decoupling D-terms.

\section{A catalogue of non-decoupling D-terms}\label{section:catalogue}
D-terms extensions of the MSSM were first explored in \cite{Batra:2003nj,Maloney:2004rc}, as they may provide a tree level enhancement
of the Higgs mass $m_h$ through a modification of the Higgs quartic
terms in the scalar potential. A higher tree-level mass requires smaller loop-level corrections to reproduce the measured Higgs mass with respect to the MSSM, with improved consequences for naturalness. The main idea is the following: the D-terms
induced by an extended gauge group diagonally breaking to the MSSM's $SU(2)_L\times U(1)_Y$ contribute to
the Higgs quartic potential. The gauge symmetry breaking is caused by the acquisition of vevs by some linking fields charged under the gauge group. The minimum of the potential is in a D-flat direction, leaving the Higgs doublets massless (at tree-level). Once the heavy linking fields are integrated out, 
the associated D-terms do not decouple in the supersymmetric
limit as soft masses for the linking fields are introduced at a scale equal or higher than the breaking
scale, remaining in the Higgs scalar potential at
lower energies. The additional non-decoupling D-terms raise the Higgs tree-level mass while introducing an effective hard SUSY breaking in the quartic scalar couplings.
For more details on the generation of non-decoupling D-terms, see \cite{Batra:2003nj,Maloney:2004rc,Bharucha:2013ela} and appendix \ref{app:generalDterms}.

Non-decoupling D-terms extensions of MSSM may arise in two- (or more) site
quiver models, for example with a single linking field $L$ 
between the
sites, in the bifundamental representation under the two gauge group
copies of $SU(2)$, as is the case in
\cite{Batra:2004vc,Delgado:2004pr,Medina:2009ey,Huo:2012tw}. Alternatively, non-decoupling D-terms are
predicted in two-site quiver models with a bifundamental and
antibifundamental pair of linking
fields $L,\tilde{L}$
\cite{Csaki:2001em,Cheng:2001an,DeSimone:2008gm,McGarrie:2010qr,Auzzi:2012dv,Craig:2012hc,d2013fitting,Bharucha:2013ela,Blum:2012ii}.
Furthermore, as quiver models are related to 
extra dimensional models through deconstruction \cite{McGarrie:2010qr,McGarrie:2011dc}, non-decoupling D-terms may also appear in this latter context (this was pointed out in
\cite{Bharucha:2013ela}).

We wish to compare here the minimisation conditions and the implications
for naturalness within the MSSM and some of
its possible quiver extensions. 
 
The gauge group of the MSSM extensions we consider is given by
$G=SU(3)_c\times G_A\times G_B$, where $G_A,\,G_B$ are copies of
$SU(2)\times U(1)$ respectively located in sites $A$ and $B$.  Regardless
of how supersymmetry is broken, mediated either by gauge, gravity or
some other effect, it is reasonable to approximate the low energy theory
of these two-site models with the MSSM supplemented by an effective
action to account for the D-terms.  This approach neglects RGE effects,
while the full implementations of the UV completions, although warranted
such as in \cite{Bharucha:2013ela}, are beyond the scope of this work.

\subsection{Minimal Supersymmetric Standard Model (MSSM)}
It is useful, in the context of the MSSM and its D-term extensions, to use the most general renormalizable scalar potential for a two Higgs-doublet model (2HDM) \cite{Gunion:2002zf}, 
\begin{align}\small
\mathcal{V}=\, &m_1^2|H_u|^2 + m_2^2|H_d|^2 +m_{12}^2 (H_uH_d +H^{\dagger}_uH^{\dagger}_d)   \nonumber\\
&+ \frac{\lambda_1}{2}|H_d|^4+\frac{\lambda_2}{2}|H_u|^4+ \lambda_3|H_u|^2|H_d|^2 +\lambda_4|H^{\dagger}_dH_u|^2+ \frac{\lambda_5}{2}[(H_u\cdot H_d)^2+ \mbox{c.c.}]\nonumber\\
&+\lambda_6|H_d|^2[(H_u\cdot H_d)+\mbox{c.c.}]+\lambda_7|H_u|^2[(H_u\cdot H_d)+\mbox{c.c.}]\,,\label{2HDMPotential}
\end{align}
with all parameters real and CP-conserving. To recover the MSSM Higgs scalar potential, we take
\begin{align}
 m_1^2&=(\left|\mu\right|^2+m^2_{H_u})\,,\quad m_2^2=(\left|\mu\right|^2+m^2_{H_d})\,,\quad m_{12}^2=B_{\mu}\,,\nonumber\\
 \lambda_1&=\lambda_2=\frac{g^2+g^{\prime\,2}}{4}\,,\quad -\lambda_3=\frac{g^2+g^{\prime\,2}}{4}\,,\quad\lambda_4=\frac{1}{2}g^2\,,\quad\lambda_5=\lambda_6=\lambda_7=0\,,\label{MSSMPotentialparameters}
\end{align}
 with $g',\,g$ respectively being the Standard Model hypercharge and the $SU(2)_L$ coupling constants.\footnote{In the following, we take $g_1$ to be $SU(5)$ GUT-normalized, such that $g_1 = g_{1,GUT}=\sqrt{5/3}g'$, $g_2=g$ and $m_Z^2=\frac{1}{4}(\frac{3}{5}g_1^2+g_2^2)(v_u^2+v_d^2)$.}


The up- and down-Higgs doublet scalar fields may be written in terms of their charged and neutral components, $H_u=(H^+_u,H^0_u),\,H_d=(H^0_d,H^-_d)$. The minimisation conditions $\frac{\partial \mathcal{V}}{\partial H_{d}^0}=0=\frac{\partial \mathcal{V}}{\partial H_{u}^0}$ should be fulfilled for the consistency of the electroweak breaking minimum of the potential. The vevs of the neutral components are defined as \cite{Djouadi:2005gj} 
\begin{equation}
 \braket{H_u^0}=\frac{v_u}{\sqrt{2}}\,, \quad \braket{H_d^0}=\frac{v_d}{\sqrt{2}}\,, \ee
\be  v^2\equiv v_u^2+v_d^2=(246 \mbox{ GeV})^2 \, ,  \quad \frac{v_u}{v_d}\equiv\frac{v \sin \beta}{v \cos \beta } = \tan \beta\, .
\end{equation}

\noindent The minimisation condition equations then read
\begin{equation}\label{TadpoleMSSM1}
 m_{H_u}^2+ |\mu|^2-B_{\mu}\cot \beta - \frac{m_Z^2}{2}\cos (2\beta)=0\,,
\end{equation}
\begin{equation}
 \label{TadpoleMSSM2}
m_{H_d}^2+ |\mu|^2-B_{\mu}\tan\beta +\frac{m_Z^2}{2}\cos (2\beta)=0\,,
\end{equation}

\noindent where $m_{H_u}$ and $m_{H_d}$ are the Higgs soft masses and $B_{\mu}$ is the MSSM $b$-term.
Taking $m_Z^2$ and $\tan \beta$ as output parameters, eqns. \eqref{TadpoleMSSM1},\eqref{TadpoleMSSM2} can be rewritten as:
\begin{align}
\sin (2\beta)&=\frac{2B_{\mu}}{m_{H_u}^2+m_{H_d}^2 + 2|\mu|^2}\,,  \\
m_Z^2 &=\frac{|m_{H_d}^2-m_{H_u}^2|}{\sqrt{1-\sin^2(2\beta)}}-m_{H_u}^2-m_{H_d}^2-2|\mu|^2\,.
\end{align}

In the MSSM, after electroweak symmetry breaking there are five physical scalar states: the two CP even neutral scalars $h$ and $H$, the CP odd neutral scalar $A^0$, and the conjugate charged Higgses $H^+,H^-$. Using the tree-level scalar potential minimised around the vevs  $v_u$ and $v_d$, one obtains the set of masses 
\begin{align}
 m^{2{\rm,\, MSSM}}_{h,H}&=\frac{1}{2}\left(m^2_{A^0} +m_Z^2 \mp \sqrt{(m_{A^0}^2 - m_Z^2)^2+ 4m_Z^2 m_{A^0}^2\sin^2 (2\beta)} \right)\,,\label{CPEvenMasses}\\
m_{A^0}^{2{\rm,\, MSSM}}&\equiv \frac{2 B_{\mu}}{\sin 2\beta }=2|\mu|^2 + m_{H_u}^2+m_{H_d}^2\,,\label{CPOddMass}\\
m_{H^{\pm}}^{2{\rm,\, MSSM}}&=m_{A^0}^2+m_{W}^2\,.\label{ChargedHMass}
\end{align}

\begin{figure}[thb!]
\begin{center}
\includegraphics[height=4cm]{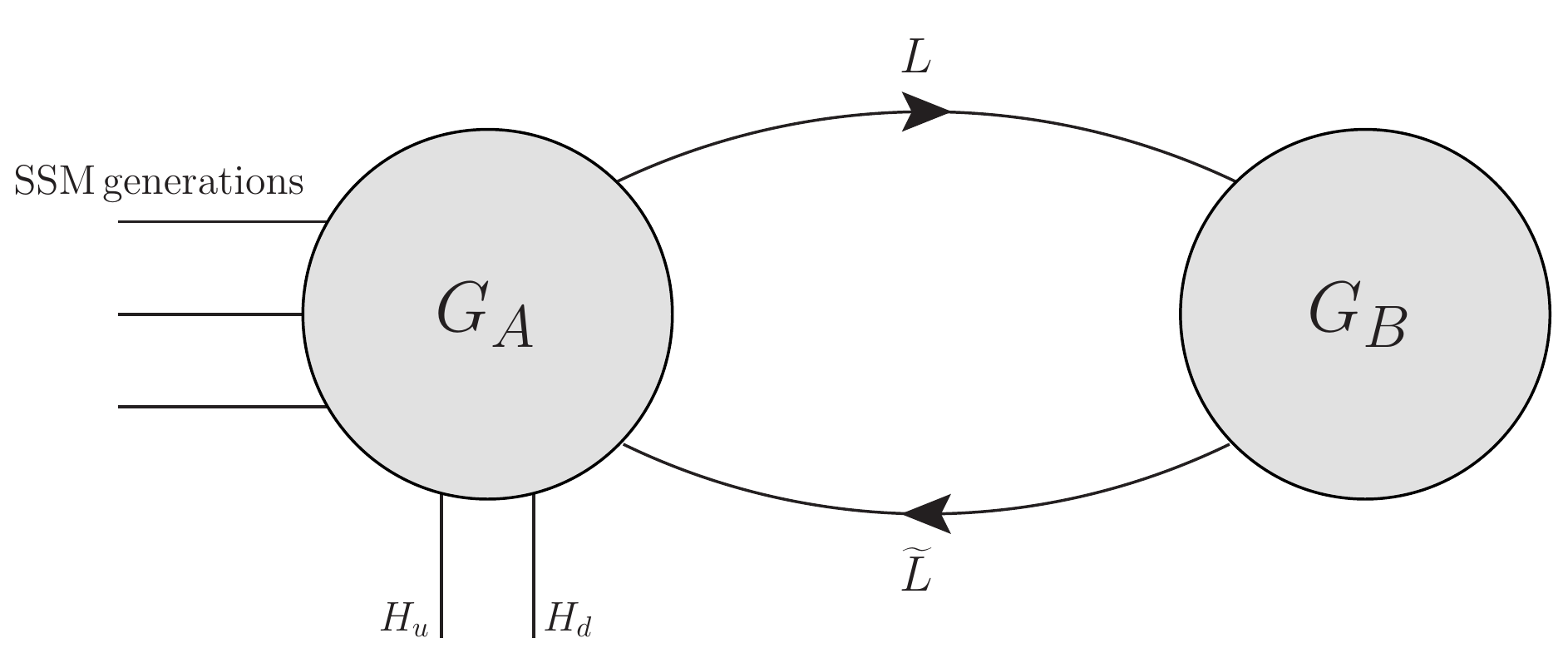}
\caption{The quiver module of the electroweak sector which leads to the vector-Higgs D-term, as in table \ref{Table:matterfieldsVeCModelA}.  The
supersymmetric standard model is on site A, the linking
fields ($L,\tilde{L}$) connect the two sites.  The singlet field ($K$) is not shown. The resulting non-decoupling vector-Higgs D-term is displayed in \refe{eq:D-terms}.} 
\label{fig:ModelA}
\end{center}
\end{figure}


The tree level Higgs mass is bounded by $m_{h,0}^2<m_Z^2\cos^22\beta$, requiring large loop corrections to reproduce the measured SM-like Higgs mass at $\sim125.5$ GeV.
The MSSM Higgs mass squared in the decoupling limit $m_{A^0}\gg m_Z$
can be approximated at one loop (with two-loop leading-log effects included) by \cite{Ellis:1991zd,Lopez:1991aw,Carena:1995bx,Haber:1996fp,Degrassi:2002fi,Drees:2004jm}
\begin{equation}
 m_{h,1}^{2{\rm,\, MSSM}}\simeq m_{z}^2\cos^2 2\beta +
\frac{3}{2\pi^2v^2}\left[m_{t,\,r}^4\left(\sqrt{m_tM_{\tilde{t}}}\right)\ln
\frac{M^2_{\tilde{t}}}{m_t^2}+m_{t,\,r}^4(M_{\tilde{t}})\frac{X^2_t}{M_{\tilde{t}}^2}\left(1-\frac{X^2_t}{12M_{\tilde{t}}^2}
\right)\right]\,,\label{Eq:1loopmhMSSM}
\end{equation}
where $m_{t,\,r}(\Lambda)$ is the running top mass at the scale $\Lambda$ and $M^2_{\tilde{t}}=m_{\tilde{t}_1}m_{\tilde{t}_2}$; $v=246$ GeV is the electroweak Higgs vev and $X_t=A_t-\mu^{\ast}\cot\beta$, with $A_t$ the stop soft SUSY-breaking trilinear coupling, which quantifies stop mixing.\footnote{In the following we assume $\mu$ to be real.} This expression assumes that the left and right soft parameters of the stops
are equal, see appendix \ref{appendix:sfermionmatrix} for the stop mixing matrices.

\subsection{Vector Higgs quiver model}
The first class of two-sites quiver models that we consider is given by the ``vector Higgs'' case, in which both the Higgs doublets of the MSSM are on the same site \cite{Batra:2003nj,Maloney:2004rc}. Depicted in figure \ref{fig:ModelA}, the case in which both the Higgs doublets and the other MSSM matter fields are on site $A$, charged under $G_A$ as described in table \ref{Table:matterfieldsVeCModelA}.
\begin{table}
\begin{center} 
\begin{tabular}{|c|c|c|c|c|} 
\hline 
Superfields & Spin 0 & Spin \(\frac{1}{2}\) & $G_A \times G_B \times SU(3)_c$ \\
\hline 
\(\hat{q}^{f}\) & \(\tilde{q}^{f}\) & \(q^{f}\) & \(({\bf 2}, \frac{1}{6}, {\bf
1}, 0, {\bf 3}) \)  \\ 
\(\hat{d}^{f}\) & \(\tilde{d}_R^{f*}\) & \(d_R^{f*}\) & \(({\bf 1}, \frac{1}{3},
{\bf 1}, 0, {\bf \overline{3}}) \)  \\ 
\(\hat{u}^{f}\) & \(\tilde{u}_R^{f*}\) & \(u_R^{f*}\) & \(({\bf 1},-\frac{2}{3},
{\bf 1}, 0, {\bf \overline{3}}) \)  \\
\(\hat{l}^{f}\) & \(\tilde{l}^{f}\) & \(l^{f}\) & \(({\bf 2},-\frac{1}{2}, {\bf
1} ,0, {\bf 1}) \) \\
\(\hat{e}^{f}\) & \(\tilde{e}_R^{f*}\) & \(e_R^{f*}\) & \(({\bf 1}, 1, {\bf 1},
0, {\bf 1}) \)  \\  \hline\hline
\(\hat{H}_d\)   & \(H_d\) & \(\tilde{H}_d\)     & \(({\bf 2},-\frac{1}{2}, {\bf
1}, 0, {\bf 1}) \)  \\ 
\(\hat{H}_u\) & \(H_u\) & \(\tilde{H}_u\)       & \(({\bf 2}, \frac{1}{2}, {\bf
1}, 0, {\bf 1}) \)  \\ 
 \hline\hline
\(\hat{L}\) & \(L\) & \(\psi_L\) & \(({\bf 2}, -\frac{1}{2}, {\bf \overline{2}},
\frac{1}{2}, {\bf 1}) \)   \\ 
\(\hat{\tilde{L}}\) & \(\tilde{L}\) & \(\psi_{\tilde{L}}\) & \(({\bf
\overline{2}}, \frac{1}{2}, {\bf 2}, -\frac{1}{2}, {\bf 1}) \)  \\
\(\hat{K}\) & \(K\) & \(\psi_{K}\) & \(({\bf 1}, 0, {\bf 1}, 0, {\bf 1}) \)  \\
\hline 
\end{tabular} \caption{The matter content of the theory that may lead to a vector-Higgs non decoupled D-term for both $SU(2)_L$ and $U(1)_Y$, with the Higgs doublets on site $A$.   $f=1,2,3$ labels the generations. The singlet $\hat{K}$ couples to the linking fields in the superpotential and it is introduced to generate a suitable scalar potential for the linking fields, see also \cite{Bharucha:2013ela}. This model is represented in figure \ref{fig:ModelA}.
\label{Table:matterfieldsVeCModelA}}
\end{center} 
\end{table}



 As outlined at the beginning of the section, after the symmetry breaking of $G_A\times G_B$ to $SU(2)_L\times U(1)_Y$, the real uneaten scalar components of the linking fields
appear in both the $A$ and $B$ site scalar D-term potential. When these components are integrated out, in the effective theory  the following relevant terms are added to the MSSM Higgs potential
\begin{equation}
 \delta \mathcal{L}=-\frac{3}{5}\frac{g_1^2\Delta_1}{8}  (H^{\dagger}_u
H_u-{H}^{\dagger}_d H_d)^2-\frac{g_2^2\Delta_2 }{8}\sum_a
(H^{\dagger}_u\sigma^a H_u+{H}^{\dagger}_d\sigma^aH_d)^2  +\ldots \,.\label{eq:D-terms}
\end{equation}
The ellipsis denote terms involving other scalars of the model as explained in appendix \ref{app:generalDterms}. 
$\Delta_1$ and $\Delta_2$, see Table \ref{Table:matterfieldsVeCModelA}, are respectively given by
\begin{equation}
 \Delta^A_1=\left(\frac{g^2_{A1}}{g^2_{B1}}\right)\frac{m_L^2}{m_{v1}^2+m_{L}^2} \
\ , \ \
\Delta^A_2=\left(\frac{g^2_{A2}}{g^2_{B2}}\right) \frac{m_L^2}{m_{v2}^2+m_{L}^2}\,, 
 \label{eq:nondecoupled}
\end{equation}
where $g_{A1},\,g_{B1}$ are the $U(1)$ couplings on site $A$ and $B$ while $g_{A2},\,g_{B2}$ are the $SU(2)$ couplings; $m_L$ is the soft mass, that we assume equal for both the linking fields $L,\,\tilde{L}$, and $m_{v_1},\,m_{v_2}$ are the masses of the heavy gauge bosons after the symmetry breaking to $SU(2)_L\times U(1)_Y$.  The relation between the MSSM gauge couplings and that of the extended gauge groups takes the form
\begin{equation}
 \cos \theta_i=\frac{g_i}{g_{Ai}}\ \ \ , \ \  \sin \theta_i=\frac{g_i}{g_{Bi}}\,.\label{thetas}
\end{equation}
To enhance the D-terms one requires $g^2_{A\,i}>g^2_{B\,i}\,$, a condition that in some cases can be problematic for perturbative unification, because, if most of the matter is charged under $G_A$, then a Landau pole may be reached below the GUT scale (see appendix \ref{sec:unification}).  If we are not concerned by coupling unification, then $\Delta_1$ and $\Delta_2$ may arise independently and in general  are not equal in value.  

\begin{figure}[t!]
\begin{center}
\includegraphics[width=0.49\textwidth]{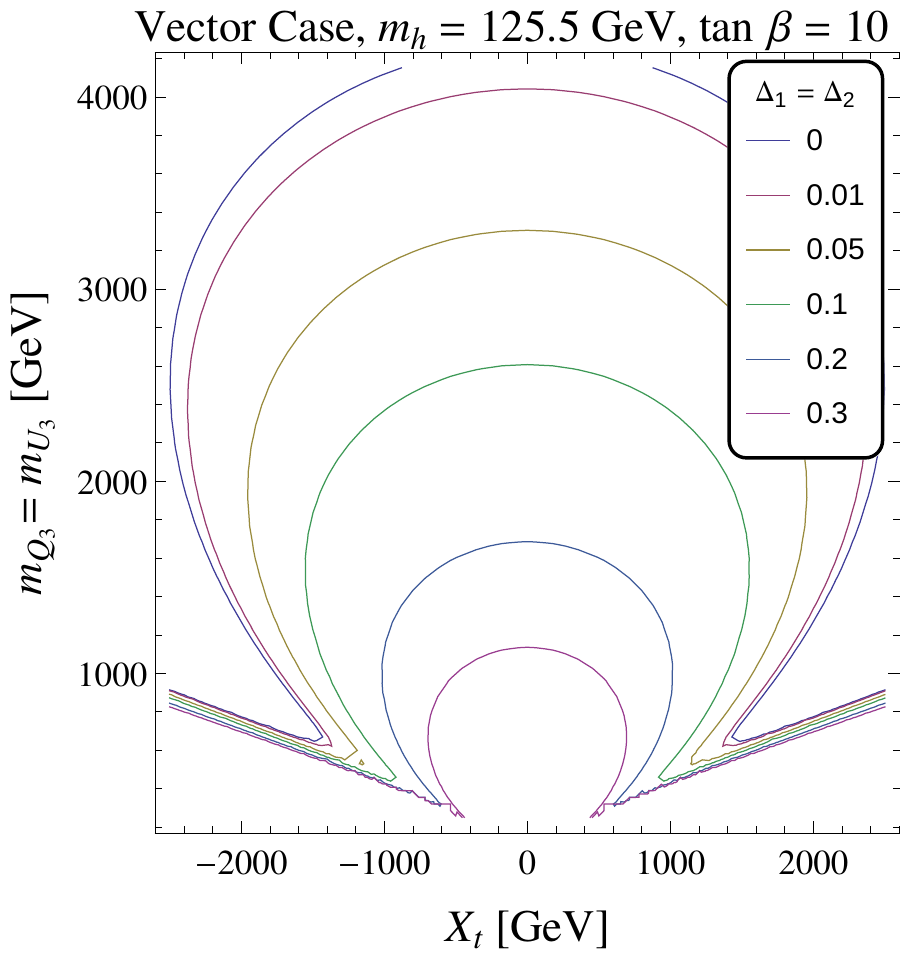}
\includegraphics[width=0.49\textwidth]{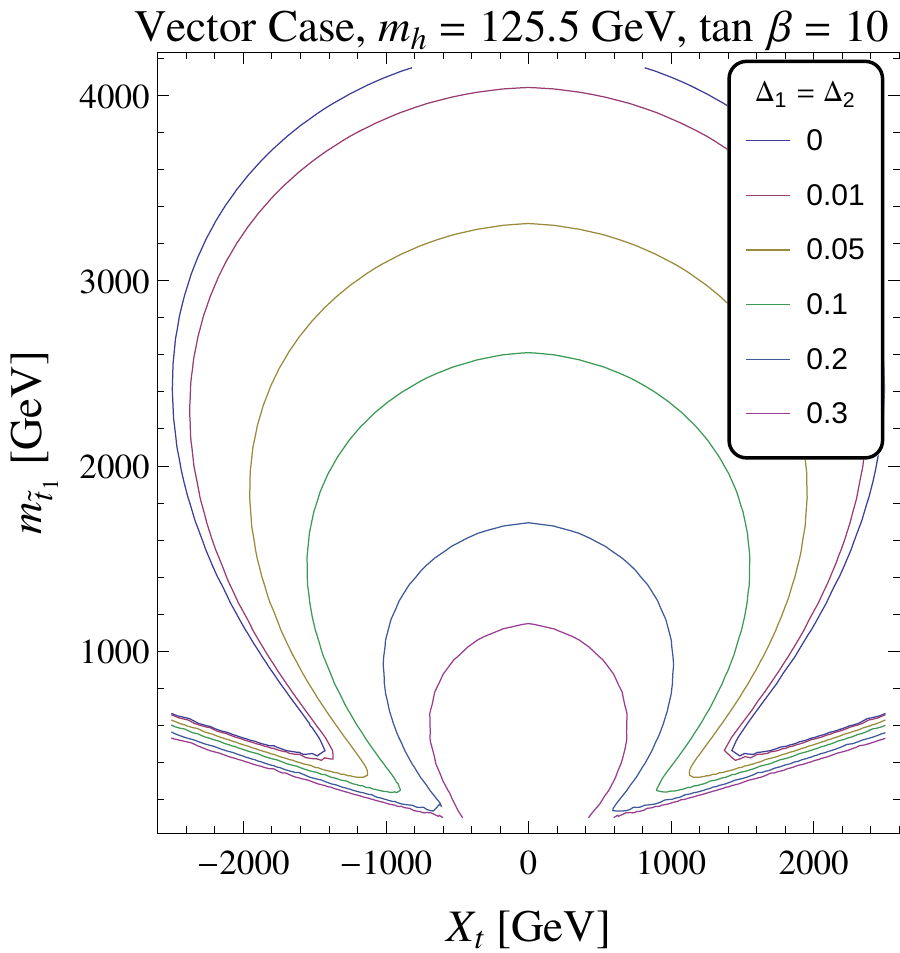}
\caption{Contours of the Higgs mass $m_h=125.5$ GeV in the ($m_{Q_3}$, $X_t$) plane [left panel] and in the ($m_{\tilde{t}_1}$, $X_t$) plane [right panel] for different values of $\Delta_1= \Delta_2$. We set $m_{Q_3}=m_{U_3}$, $\tan\beta=10$. The one-loop Higgs mass with tree-level D-terms corrections $m_{h,\,1}$ is plotted.} 
\label{VectorDeltaVSMstop1}
\end{center}
\end{figure}


For the vector Higgs extension of the MSSM, the minimisation conditions are given by
\begin{equation}
 \label{TadpoleVector1}
m_{H_u}^2+ |\mu|^2-B_{\mu}\cot \beta - \frac{m_Z^2+m^2_{\Delta}}{2}\cos (2\beta)=0,
\end{equation}
\begin{equation}
 \label{TadpoleVector2}
m_{H_d}^2+ |\mu|^2-B_{\mu}\tan\beta +\frac{m_Z^2+m^2_{\Delta}}{2}\cos (2\beta)=0,
\end{equation}
where we defined $4m^2_{\Delta}=(\frac{3}{5}g_1^2 \Delta_1 + g_2^2\Delta_2)v^2$.
Eqs. \eqref{TadpoleVector1},\eqref{TadpoleVector2} solved for $m_Z^2$ and $\tan \beta$ read
\begin{align}
\sin (2\beta)&=\frac{2B_{\mu}}{m_{H_u}^2+m_{H_d}^2 + 2|\mu|^2},  \\
m_Z^2+ m_{\Delta}^2 &=\frac{|m_{H_d}^2-m_{H_u}^2|}{\sqrt{1-\sin^2(2\beta)}}-m_{H_u}^2-m_{H_d}^2-2|\mu|^2.
\end{align}

The tree-level Higgs masses are found simply by replacing $m_Z^2\rightarrow m_Z^2+m_{\Delta}^2$ and $m_W^2\rightarrow m_W^2(1+\Delta_2)$ in \eqref{CPEvenMasses}-\eqref{ChargedHMass}.

The non-decoupling D-terms contribution causes a shift in the tree level Higgs mass squared $m^2_{h,0}$ which results in
\begin{equation}
  m^{2{\rm,\, vec} }_{h,0}= \left[m_Z^2+\left(\frac{\frac{3}{5}g^2_1\Delta_1+g^2_2\Delta_2}{4}\right)v^2\right]\cos^2 2\beta\,.
\label{eq:mh1QEW}
\end{equation}
In the following we will consider $\Delta_1$ equal to $\Delta_2$ and we will simply refer to as $\Delta$.
The effect of the tree-level shift can significantly reduce fine-tuning in the top-stop sector and allows for a reduced average stop mass. This can be seen in fig. \ref{VectorDeltaVSMstop1} (similarly to \cite{Blum:2012ii}), where we plot in the ($m_{Q_3},X_t$) and ($m_{\tilde{t}_1} ,X_t$) planes the Higgs mass from eq. \eqref{Eq:1loopmhMSSM} with the tree-level D-terms corrections from eq. \eqref{eq:mh1QEW}, for different values of $\Delta$. While at the MSSM limit $\Delta=0$, for $X_t=0$ GeV, we need $m_{\tilde{t}_1}\simeq4$ TeV to reproduce the correct Higgs, at $\Delta=0.3$ this is possible with $m_{\tilde{t}_1}\simeq1$ TeV. One can also note that 
the value of the maximal mixing scenario (the sharply acute concave kink in the contours for $|A_t|\simeq\sqrt{6}M_s$) can further allow for a significantly smaller $X_t$ for increasing $\Delta$.

Discussing the expected order of the size of these D-terms one can observe that with $\Delta\sim \mathcal{O}(1)$ the tree-level Higgs mass would already be sufficiently large to account for the observed 125 GeV Higgs mass. In \cite{Blum:2012ii} was shown that demanding fine tuning no worse than 1/10 together with light stops one would expect $\Delta\gtrsim0.5$. On the other hand, in \cite{Bharucha:2013ela} it was found that $\mathcal{O}(0.1)$ $\Delta$ is 
more easily obtainable and preferable if to accommodate perturbative unification (see also appendix \ref{sec:unification}). As these $\mathcal{O}(0.1)$ $\Delta$ can still have a noticeable effect on the Higgs mass but may have a less easily observable deviation from the MSSM, we study here the degree to which their effects can be determined at the LHC and ILC.

\begin{figure}[t!]
\begin{center}	
\includegraphics[width=0.49\textwidth]{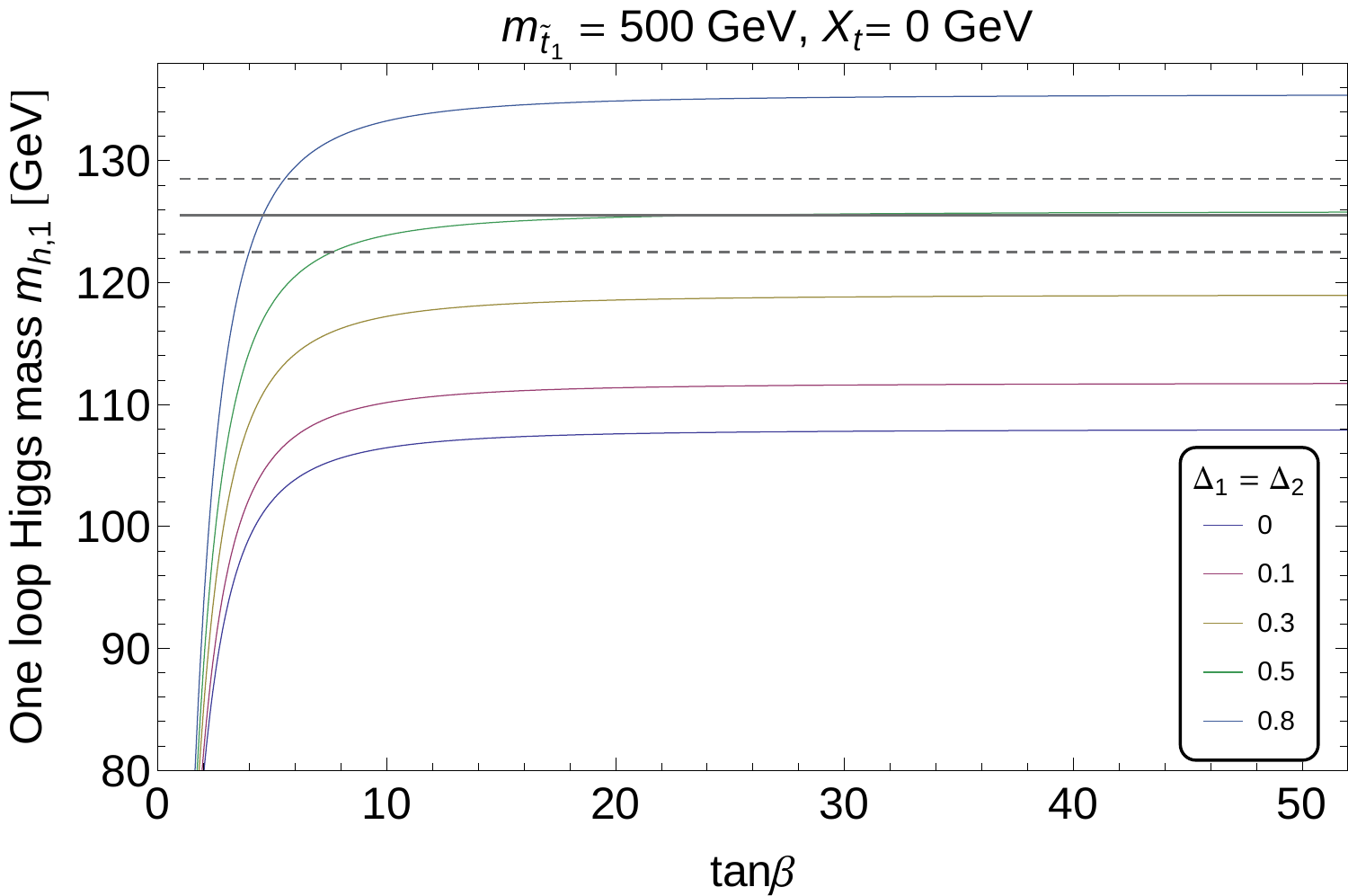}
\includegraphics[width=0.49\textwidth]{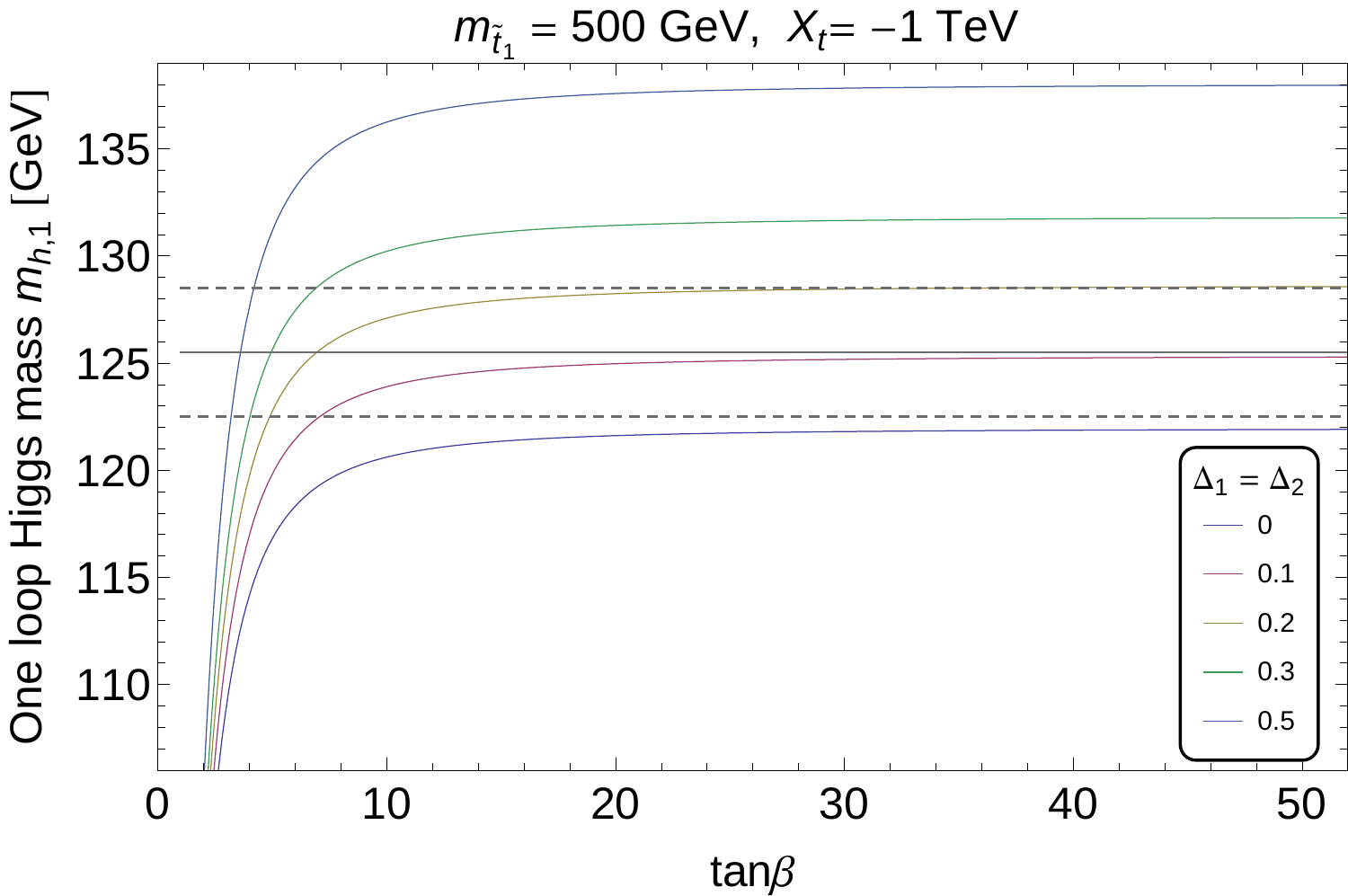}
\caption{One-loop Higgs mass $m_{h, 1}$ with tree-level D-terms corrections vs $\tan \beta$ for different values of $\Delta_1=\Delta_2$ with $X_t=0$ [left panel] and $X_t=-1$ TeV [right panel], and with $m_{\tilde{t}_1}=500$ GeV. For comparison, $125.5\pm 3$ GeV grid lines are plotted.} 
\label{tanbetaplots}
\end{center}
\end{figure}

\begin{figure}[t!]
\begin{center}
\includegraphics[width=0.49\textwidth]{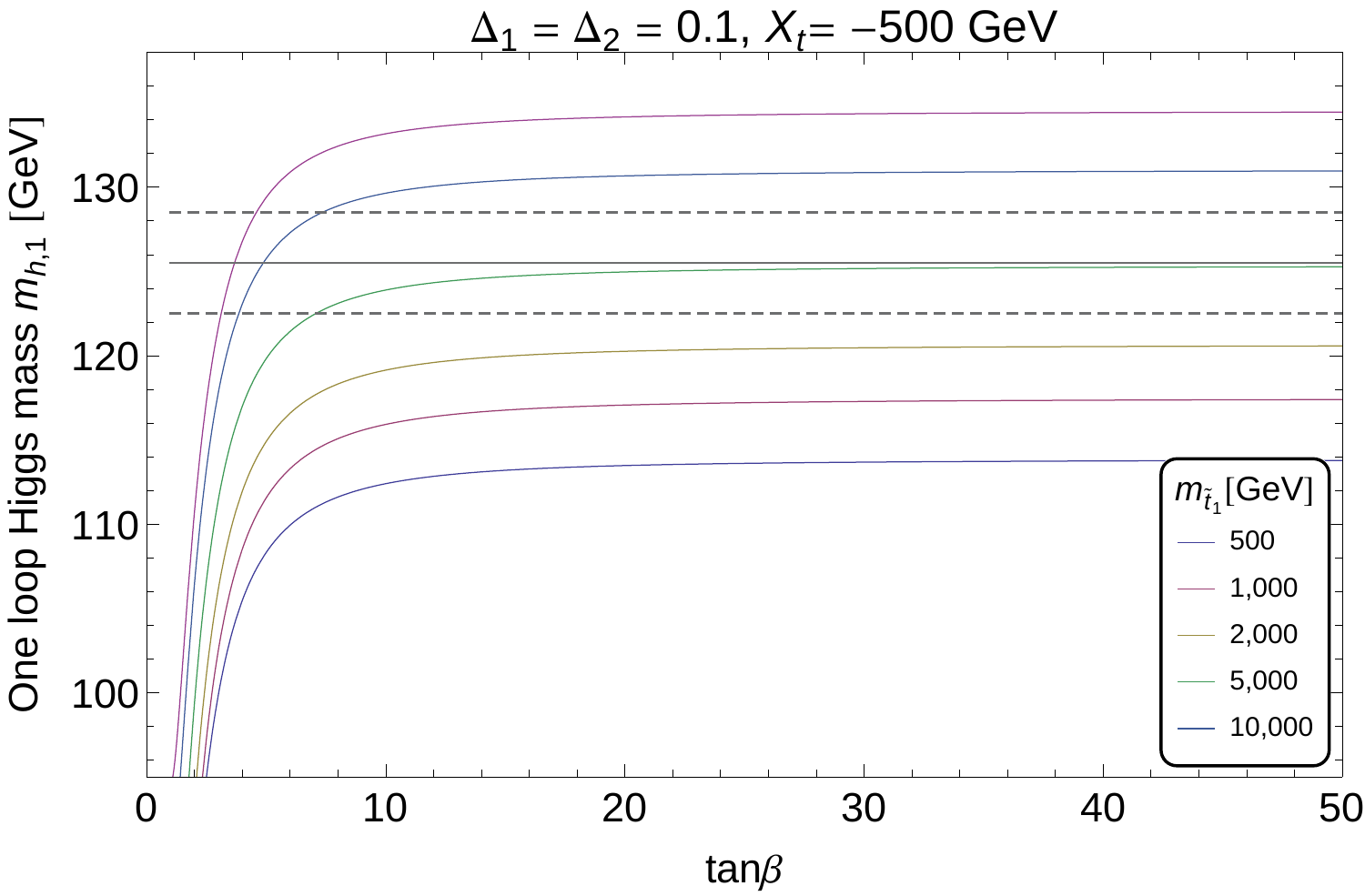}
\includegraphics[width=0.49\textwidth]{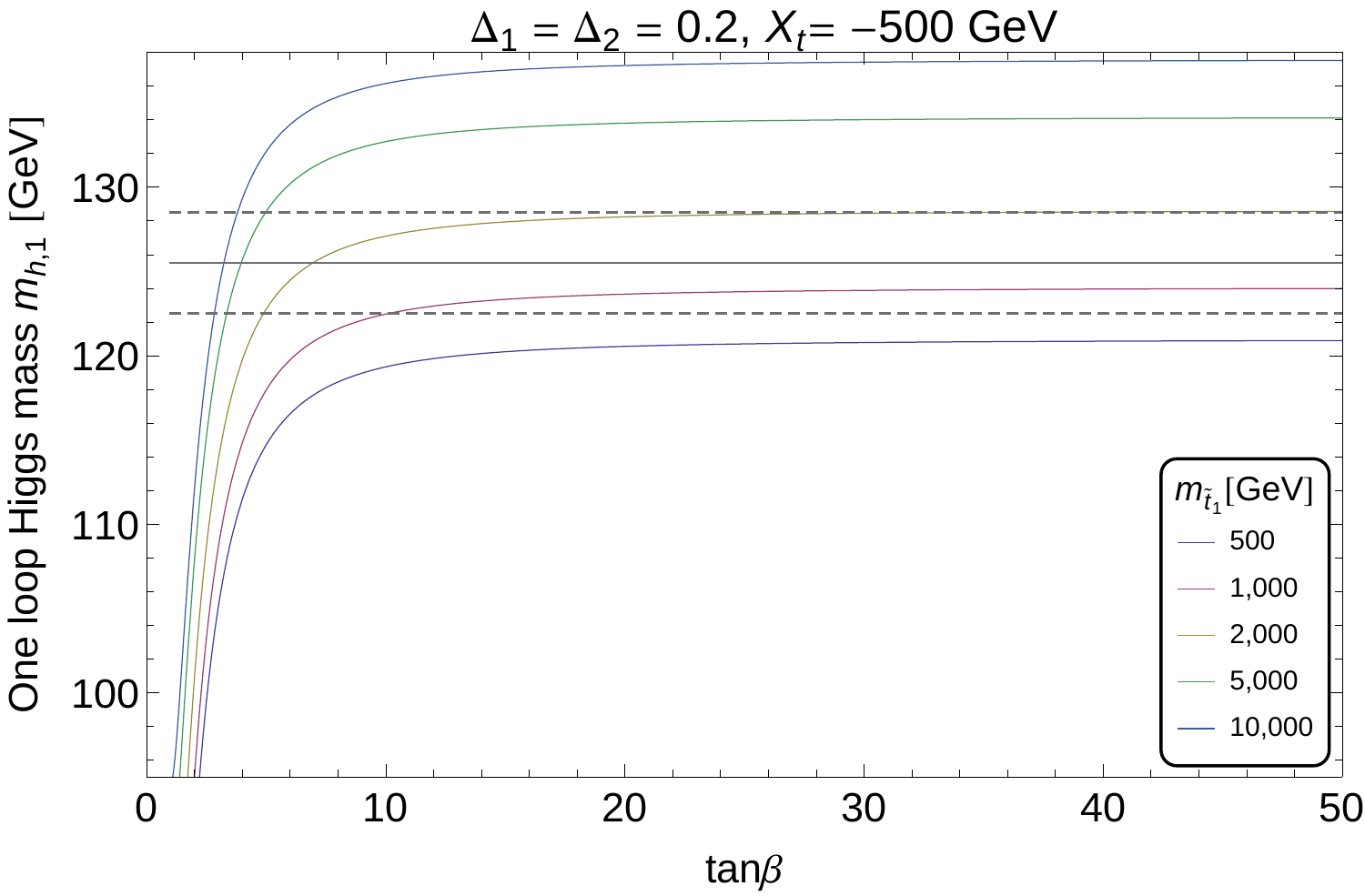}
\caption{One-loop Higgs mass $m_{h, 1}$ with tree-level D-terms corrections vs $\tan \beta$ for different values of $m_{\tilde{t}_1}$ with $\Delta_1=\Delta_2=0.1$ [left panel] and $\Delta_1=\Delta_2=0.2$ [right panel], with $X_t=-500$ GeV.  For comparison, $125.5\pm 3$ GeV grid lines are plotted. } 
\label{tanbetaplotsstop}
\end{center}
\end{figure}

 In fig.~\ref{tanbetaplots} and fig.~\ref{tanbetaplotsstop} one can see how enhancements due to the non-decoupling D-terms arise significantly for $\tan\beta\in[1,10]$, where it is harder to reproduce $m_h=125.5$ GeV, and stabilises for $\tan\beta\gtrsim10$. Such results are similarly reproduced using the RG-evolution approach as in \cite{Bharucha:2013ela}.  In particular in figure \ref{tanbetaplotsstop} it is evident that for an increasing value of $\Delta$, a lower $m_{\tilde{t}_1}$ is required to get $m_h=125.5$ GeV, especially compared to the MSSM limit of $\Delta=0$. 

\begin{figure}[t!]
\begin{center}
\includegraphics[width=0.49\textwidth]{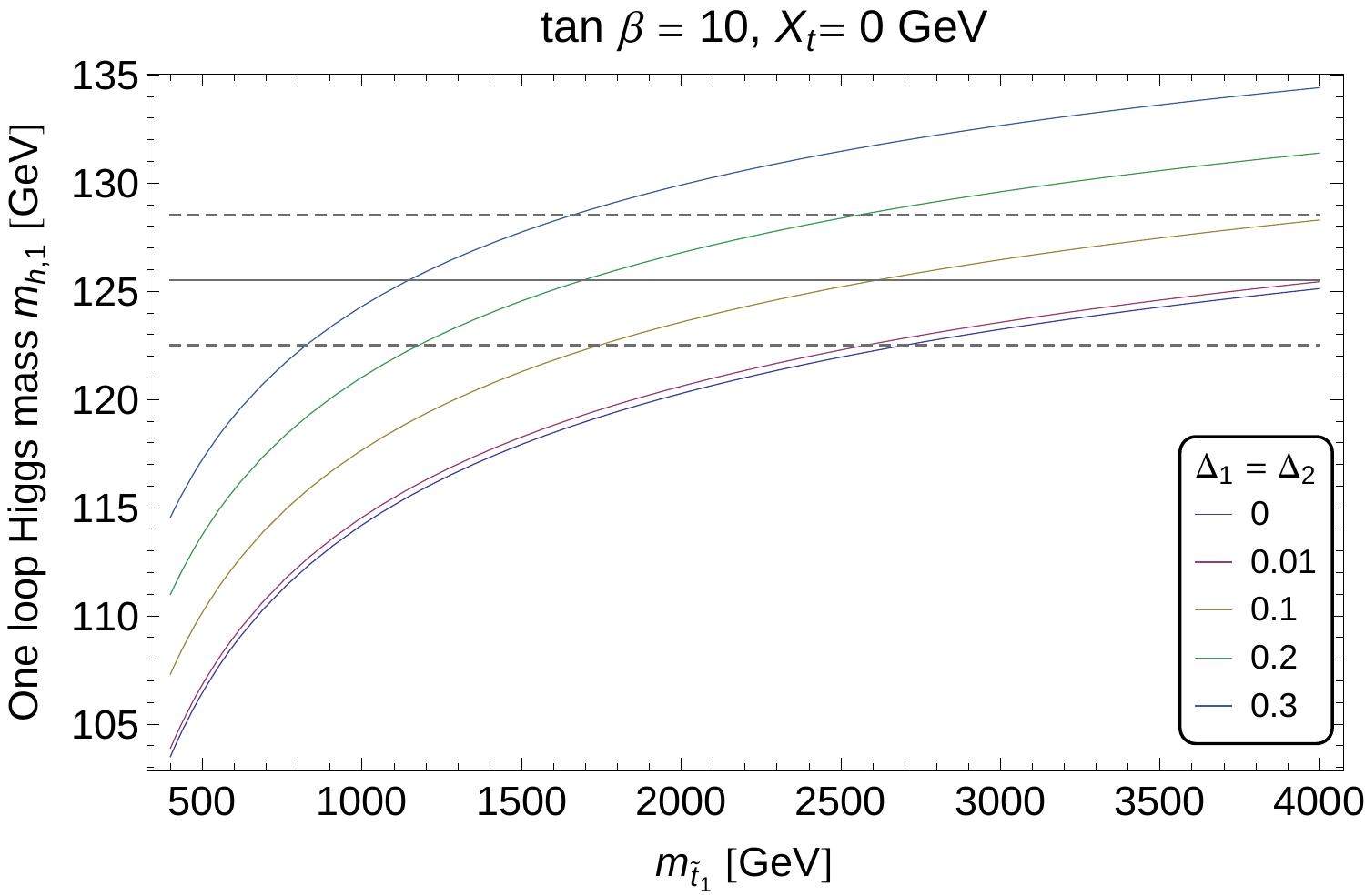}
\includegraphics[width=0.49\textwidth]{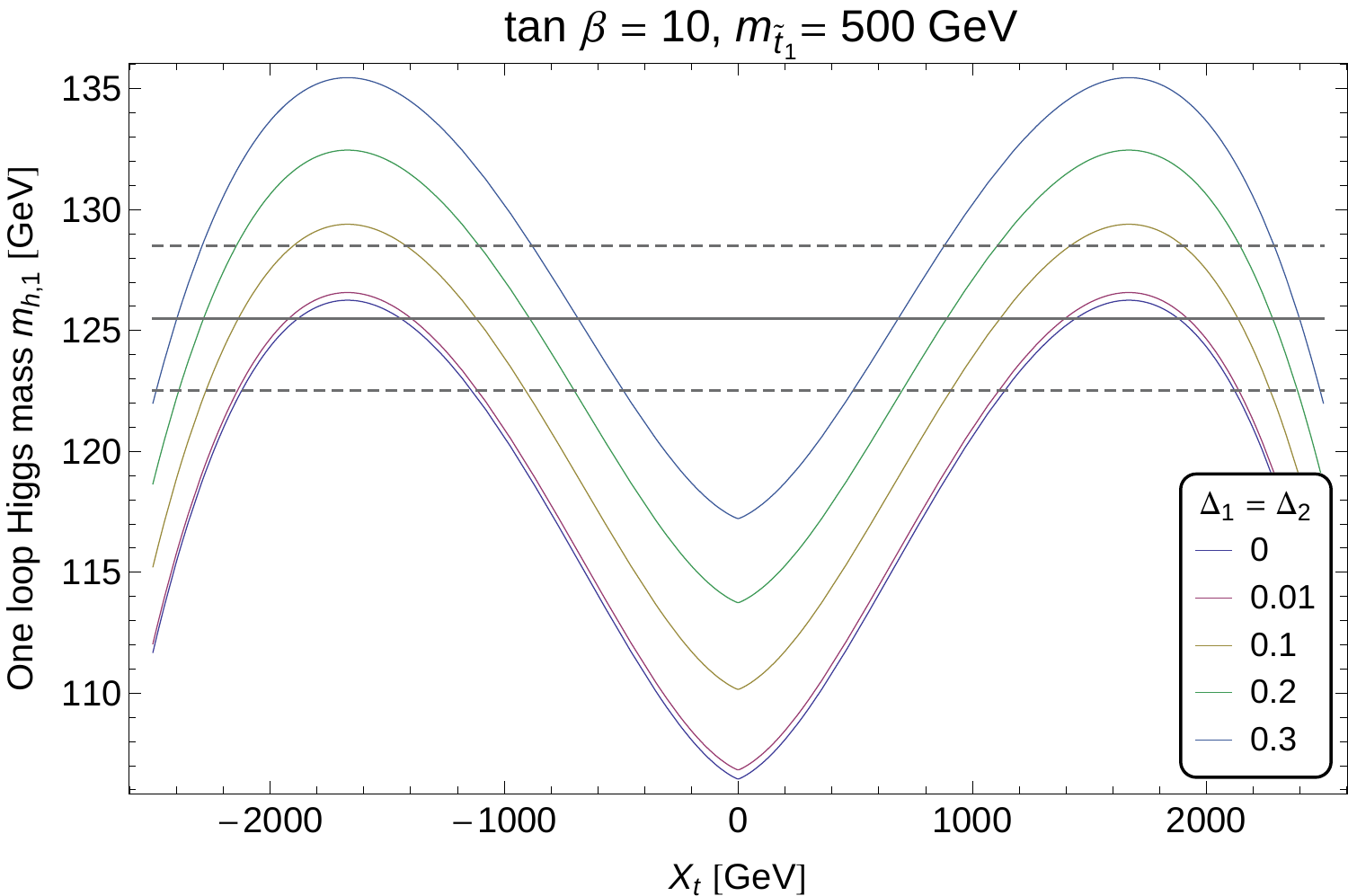}
\caption{One-loop Higgs mass $m_{h, 1}$ with tree-level D-terms corrections for different values of $\Delta_1=\Delta_2$, with $\tan \beta =10$ and $125.5\pm 3$ GeV grid lines plotted for comparison. On the left panel, $m_{h, 1}$ vs $m_{\tilde{t}_1}$ with $X_t=0$ GeV; on the right panel, $m_{h, 1}$ vs $X_t$ with $m_{\tilde{t}_1}=500$ GeV.
\label{mhversus_stop1_OR_Xt}}
\end{center}
\end{figure}

In the left panel of fig. \ref{mhversus_stop1_OR_Xt} we can see that to have null mixing $X_t=0$ GeV with $\tan\beta=10$, $m_{{\tilde{t} }_{1}}$ has to be in the 1-4 TeV range for $\Delta\in[0.01,0.3]$. On the right panel we see that for the same values of  $\Delta$  with a lower stop mass ($m_{{\tilde{t} }_{1}}$ $\sim 500$ GeV) still a $|X_t|\sim 1$ TeV is required, with negative values of $X_t$ preferred by theory due to RGE effects, which makes $A_t=X_t+\mu\cot\beta$ run negative.  In summary, whilst the maximal mixing scenario is favoured, it is now much more achievable, due to the D-term effects, for smaller values of $A_t$, and even allows sub 2 TeV stops the for the null or small mixing scenario, when $\Delta\ge 0.3$.

The vector Higgs D-term extensions of the MSSM may feature different generations of matter located on different sites, for example having the first two generation matters on site $B$ \cite{Batra:2004vc,Delgado:2004pr}, while typically the 3rd generation is on the same site as $H_u$ since the stop mixing parameter $X_t$ helps to trigger EWSB.
In alternative version of the vector-Higgs D-terms,  the Higgses are both on site $B$. The corresponding D-terms are now given by \refe{eq:D-terms} with $\Delta_1$ and $\Delta_2$ respectively equal to 
\begin{equation}
 \Delta^B_1=\left(\frac{g^2_{B1}}{g^2_{A1}}\right)\frac{m_L^2}{m_{v1}^2+m_{L}^2} \
\ , \ \
\Delta^B_2=\left(\frac{g^2_{B2}}{g^2_{A2}}\right) \frac{m_L^2}{m_{v2}^2+m_{L}^2}. 
 \label{eq:nondecoupledB}
\end{equation}

Notice that the role of the gauge couplings are reversed with respect to model A, with $g^2_{B1}>g^2_{A1}$. This can result in an easier perturbative unification if more matter is on site A than site B, although this can generate problems with EWSB and also separately naturalness, depending on where the source of supersymmetry breaking is introduced, in the context of supersymmetry breaking, for instance Non-universal UV Higgs soft masses may be required to trigger EWSB at low scales.

\begin{figure}[t!]
\begin{center}
\includegraphics[height=4cm]{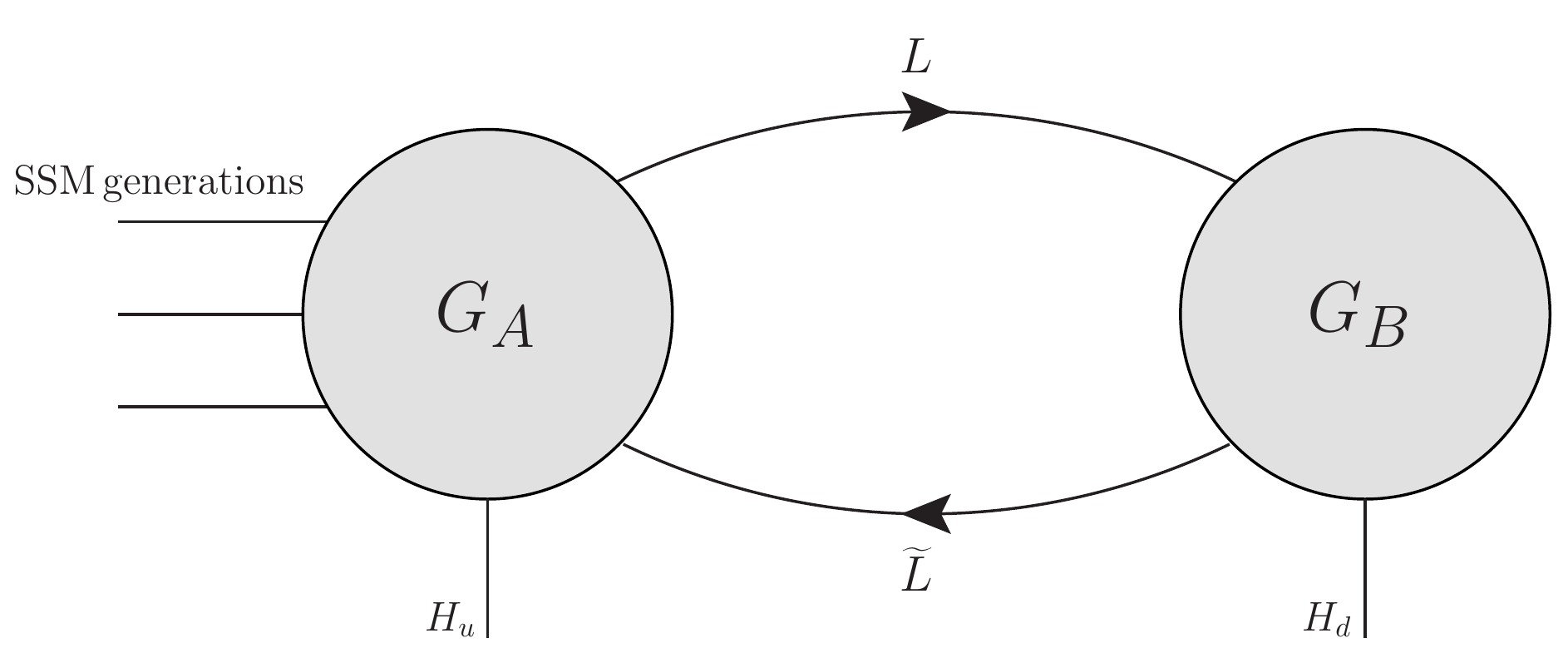}
\caption{The quiver module of the electroweak sector for the chiral Higgs case. The resulting chiral-Higgs like non-decoupling D-term is reported in \refe{eq:D-termsC} and whose matter content is in table \ref{Table:matterfieldsChirModelC}. The model requires additional fields carrying Higgs-like charges, such as in figure \ref{fig:UVofModelC}, or leptons multiplets on site B instead of A, for anomaly cancellation. } 
\label{fig:ModelC}
\end{center}
\end{figure}

\subsubsection{Additional fine-tuning and the Higgs mass}\label{FTdterms}
We should also consider the effect on naturalness of the explicit
breaking of supersymmetry from the non-decoupled D-terms in the EFT we
restrict to. Using a cut-off, the non-decoupled D-terms in the vector case will lead to a
quadratic divergence that contributes to the Higgs mass counterterm at
one loop \cite{Maloney:2004rc,Blum:2012ii}:
\begin{equation}
   \delta m_{h,1}^{2,\text{vec}}= \left(\frac{\alpha\frac{3}{5}g^2_1\Delta_1+\beta
g^2_2\Delta_2}{4}\right)\frac{M^2}{16\pi^2}\,,  \label{eq:div}
\end{equation}
where $\alpha,\beta$ are determined by the precise matter content that
appears in the non-decoupling D-term, each generating a one loop
contribution (see section \ref{app:generalDterms}), and
$M^2=m_L^2$. Such an effect may arise both in the Higgs tadpole
equations and in the one-loop Higgs self energies. In a supersymmetric
theory that is only softly broken all quadratic divergences cancel
exactly at all orders in perturbation theory. In this case, \refe{eq:div}
gives an additional contribution depending on the size of $M^2$, that
should then not be too large in order not to have too much additional
fine tuning. This fine tuning $F$, may be quantified as
\begin{equation}
\frac{\delta m_{h,1}^{2,\text{vec}}}{m_h^2}=1/F\,.
\end{equation}
In either case we have assumed in this paper, as in
\cite{Maloney:2004rc,Blum:2012ii}, that $M^2$ is small enough such that
this contribution is neglected. It is interesting to consider the
inclusion of these terms if one considers larger values of $M^2$ such as
might arise from future $Z'$ exclusions.

\subsection{Chiral Higgs quiver model}\label{subsec:ChiralCase}

Another possible two-sites quiver is the chiral Higgs case \cite{Randalltalk,Craig:2012bs}, in which the two MSSM Higgs doublets are on two alternate sites. This is pictured in figure \ref{fig:ModelC} and in table \ref{Table:matterfieldsChirModelC}, in which the up-type Higgs double $H_u$ and the third generation of matter are on site $A$, while the down-type Higgs double $H_d$ and the first two generations of matter are on site $B$.
\begin{table}
\begin{center} 
\begin{tabular}{|c|c|c|c|c|} 
\hline \hline 
Superfields & Spin 0 & Spin \(\frac{1}{2}\) & $G_A \times G_B \times SU(3)_c$ \\
\hline 
\(\hat{q}^{f}\) & \(\tilde{q}^{f}\) & \(q^{f}\) & \(({\bf 2}, \frac{1}{6}, {\bf
1}, 0, {\bf 3}) \)  \\
\(\hat{d}^{f}\) & \(\tilde{d}_R^{f*}\) & \(d_R^{f*}\) & \(({\bf 1}, \frac{1}{3},
{\bf 1}, 0, {\bf \overline{3}}) \)  \\ 
\(\hat{u}^{f}\) & \(\tilde{u}_R^{f*}\) & \(u_R^{f*}\) & \(({\bf 1},-\frac{2}{3},
{\bf 1}, 0, {\bf \overline{3}}) \)  \\
\(\hat{l}^{f}\) & \(\tilde{l}^{f}\) & \(l^{f}\) & \(({\bf 2},-\frac{1}{2}, {\bf
1} ,0, {\bf 1}) \) \\
\(\hat{e}^{f}\) & \(\tilde{e}_R^{f*}\) & \(e_R^{f*}\) & \(({\bf 1}, 1, {\bf 1},
0, {\bf 1}) \)  \\  \hline\hline
\(\hat{H}_u\) & \(H_u\) & \(\tilde{H}_u\)       & \(({\bf 2}, \frac{1}{2}, {\bf
1}, 0, {\bf 1}) \)  \\ 
\(\hat{H}_d\)   & \(H_d\) & \(\tilde{H}_d\)     & \(( {\bf
1}, 0, {\bf 2},-\frac{1}{2},{\bf 1}) \)  \\ 
 \hline\hline
\(\hat{L}\) & \(L\) & \(\psi_L\) & \(({\bf 2}, -\frac{1}{2}, {\bf \overline{2}},
\frac{1}{2}, {\bf 1}) \)   \\ 
\(\hat{\tilde{L}}\) & \(\tilde{L}\) & \(\psi_{\tilde{L}}\) & \(({\bf
\overline{2}}, \frac{1}{2}, {\bf 2}, -\frac{1}{2}, {\bf 1}) \)  \\
\(\hat{K}\) & \(K\) & \(\psi_{K}\) & \(({\bf 1}, 0, {\bf 1}, 0, {\bf 1}) \)  \\
\hline \hline
\end{tabular} \caption{The matter content of a quiver model that may lead to the Chiral Higgs case and the D-term enhancement of \refe{eq:D-termsC}. This is pictured in figure \ref{fig:ModelC}. The model requires additional fields carrying Higgs-like charges, such as in figure \ref{fig:UVofModelC}, or leptons multiplets on site B instead of A, for anomaly cancellation. \label{Table:matterfieldsChirModelC}}
\end{center} 
\end{table}
The chiral Higgs case may be quite naturally achieved from a four Higgs doublet model such as that in appendix \ref{subsec:4HDM} (figure \ref{fig:UVofModelC}), in which each site has two Higgs doublets and then at lower energies a Higgs doublet for each site are integrated out, resulting in a two Higgs doublet model.  In the chiral Higgs model the non-decoupling D-terms that are added to the scalar potential of the MSSM, at low energies, are given by 

\begin{equation}
 \delta \mathcal{L}=-\frac{3}{5}\frac{g_1^2\Omega_1}{8}  ( \xi_1 H^{\dagger}_u
H_u+\frac{1}{\xi_1}{H}^{\dagger}_d H_d)^2-\frac{g_2^2\Omega_2 }{8}\sum_a
(\xi_2 H^{\dagger}_u\sigma^a H_u-\frac{1}{\xi_2}{H}^{\dagger}_d\sigma^aH_d)^2+\dots\,\,\mbox{.}\label{eq:D-termsC}
\end{equation}
The ellipsis represent terms involving other scalar particles as reported in appendix \ref{app:generalDterms}, while
\begin{equation}
 \xi_i=\frac{g_{Ai}}{g_{Bi}}\,, \quad
\Omega_1=\frac{m_L^2}{m_{v1}^2+m_{L}^2} \, , \quad\Omega_2=\frac{m_L^2}{m_{v2}^2+m_{L}^2}\,. 
 \label{eq:nondecoupledC}
\end{equation}
The minimisation conditions now take the form 
\begin{equation}
 m_{H_u}^2+ |\mu|^2-B_{\mu}\cot \beta - \frac{m_Z^2}{2}\cos (2\beta)+m^2_{\Omega}\cos^2\beta+C=0\,,\label{TadpoleChiral1}
\end{equation}
\begin{equation}
 m_{H_d}^2+ |\mu|^2-B_{\mu}\tan\beta +\frac{m_Z^2}{2}\cos (2\beta)+m^2_{\Omega}\sin^2\beta+D=0\,,  \label{TadpoleChiral2}
\end{equation}
where
\begin{equation}
 m^2_{\Omega}=\frac{v^2}{8}\sum_{i=1,2}k_i g_i^2\Omega_i\,,
\end{equation}
with $k_i=(3/5,1)$
and
\begin{equation}
 C=\frac{v^2}{8}\sum_{i=1,2}k_i g_i^2\Omega_i\xi^2_i\sin^2\beta \ \ , \ \  D=\frac{v^2}{8}\sum_{i=1,2}k_ig_i^2\Omega_i\frac{\cos^2\beta}{\xi^2_i}\, .
\end{equation}
Eqs. \eqref{TadpoleChiral1},\eqref{TadpoleChiral2} then give
\begin{align}
\sin (2\beta)&=\frac{2B_{\mu}}{m_{H_u}^2+m_{H_d}^2 + 2|\mu|^2+C+D+m^2_{\Omega} }\,,\\
&\nonumber\\
m_Z^2
&=\frac{2}{1-\tan^2\beta}\left[(C+m_{H_u}^2)\tan^2\beta-(D+m_{H_d}^2)\right]-2|\mu|^2\,.\label{MZChiral_CD}
\end{align}

\begin{figure}[t!]
\begin{center}
\includegraphics[width=0.49\textwidth]{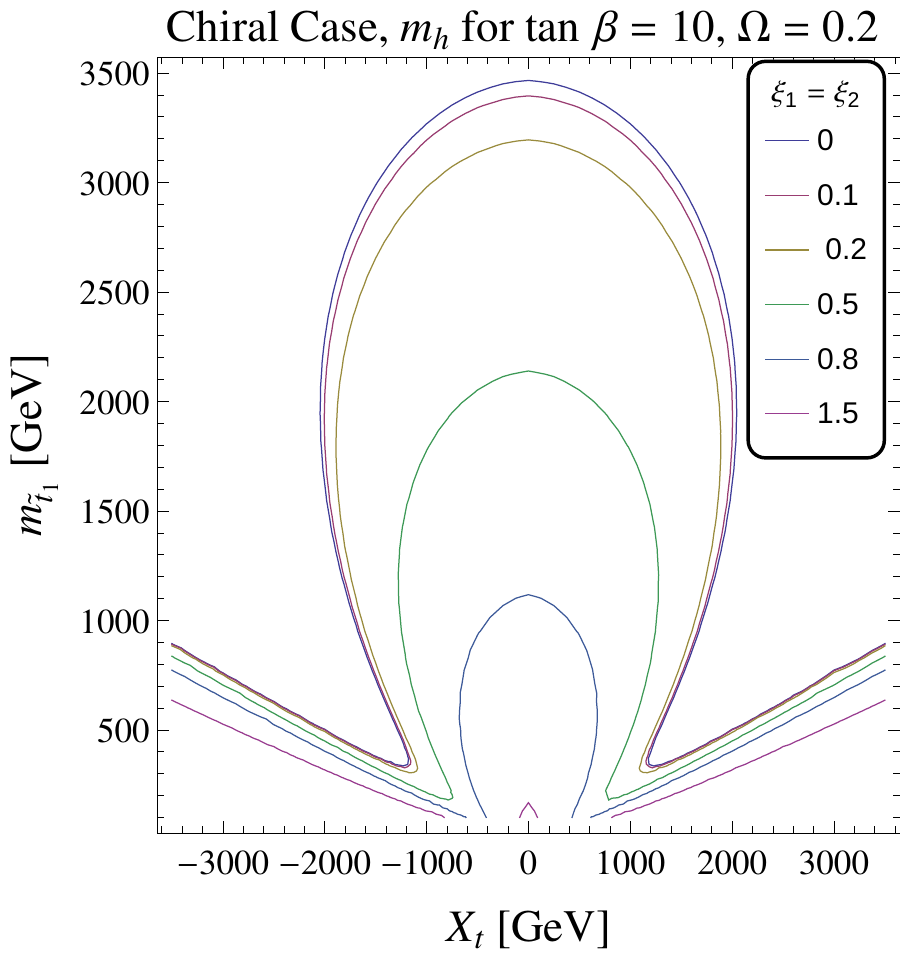}
\includegraphics[width=0.49\textwidth]{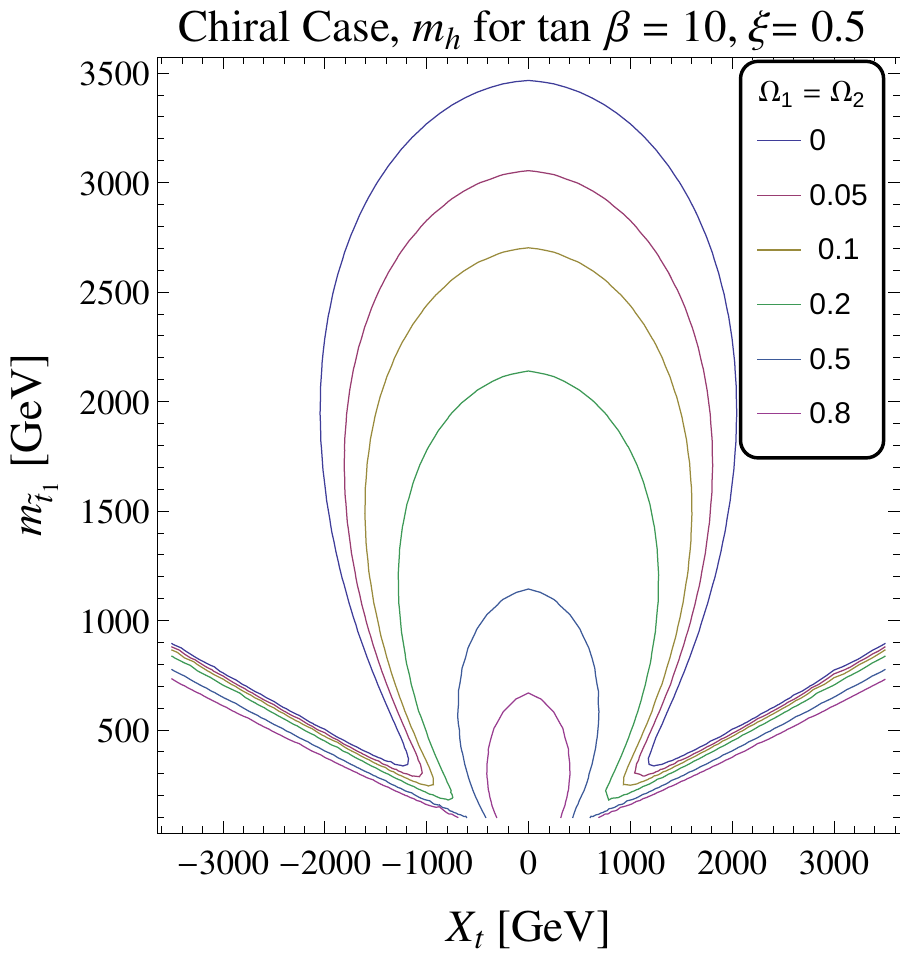}
\caption{Contours of the Higgs mass $m_h=125.5$ GeV in the ($m_{\tilde{t}_1}$, $X_t$) plane for different values of $\xi$ [left panel] and $\Omega$ [right panel], with  $m_{Q_3}=m_{U_3}$, $\tan\beta=10$. The one-loop Higgs mass with tree-level D-terms corrections $m_{h,\,1}$ is plotted.} 
\label{ChiralXiMstop1}
\end{center}
\end{figure}

\begin{figure}[t!]
\begin{center}
\includegraphics[width=0.49\textwidth]{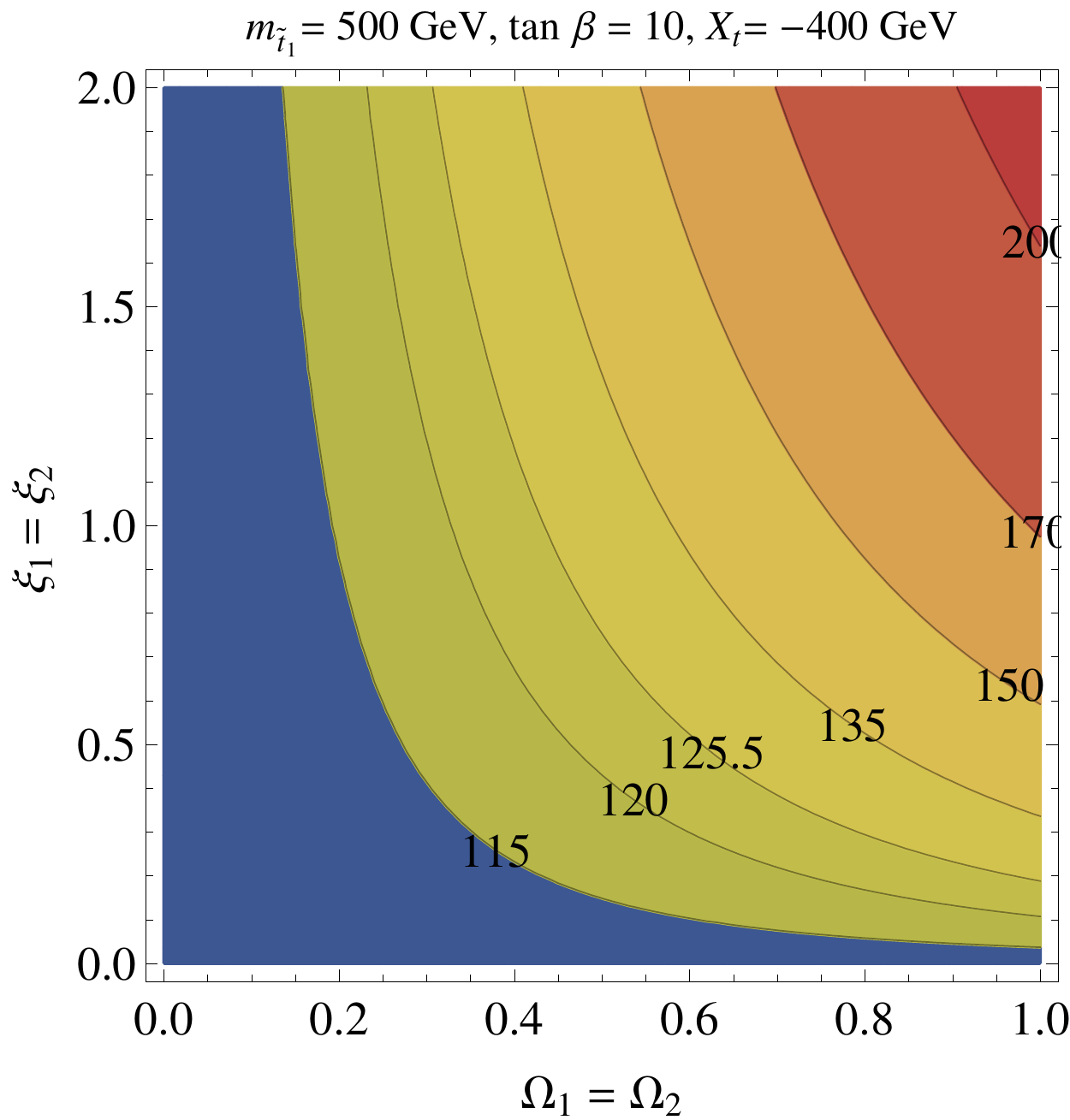}
\includegraphics[width=0.49\textwidth]{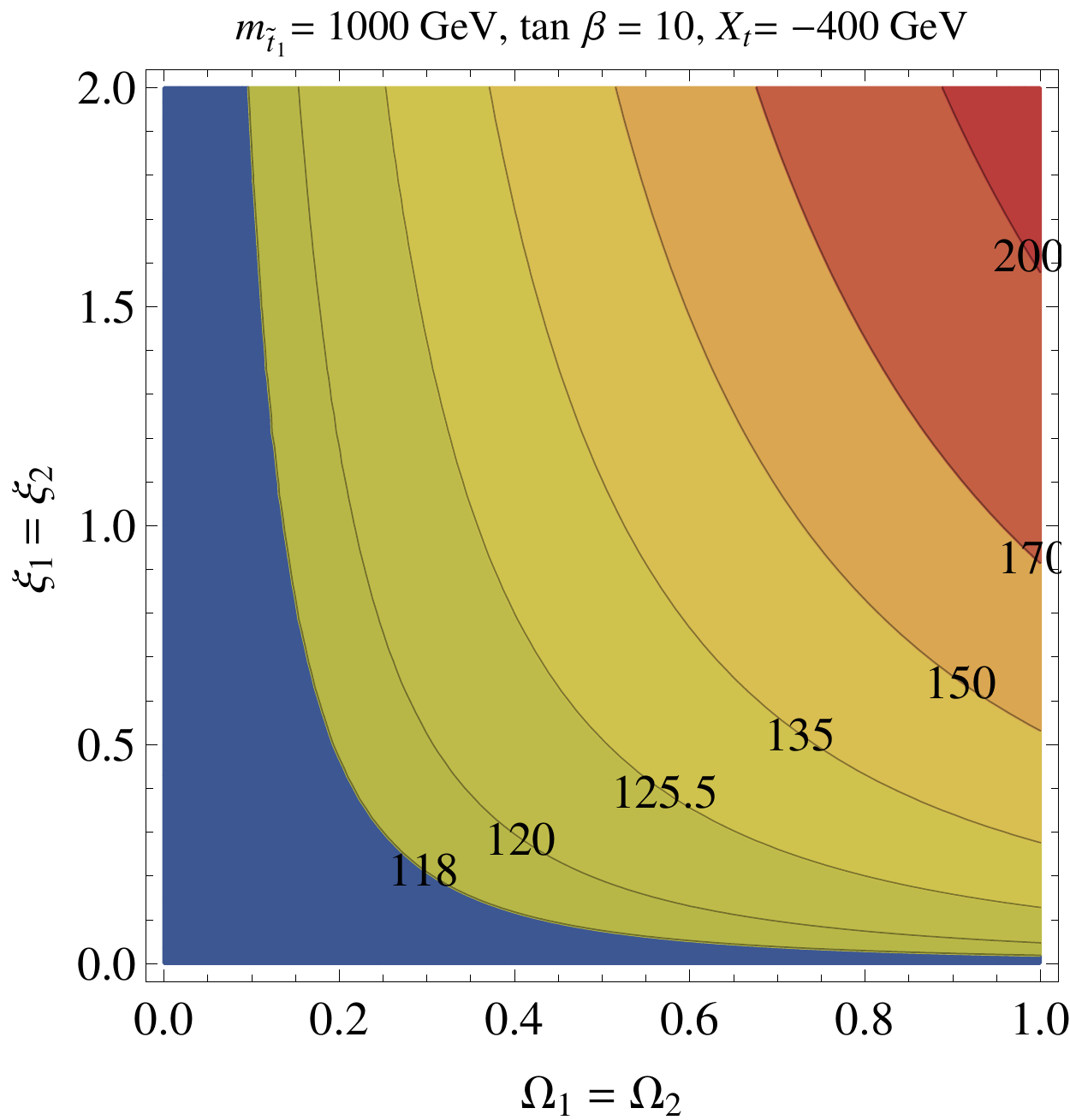}
\caption{The Higgs mass in the ($\xi_1=\xi_2$, $\Omega_1=\Omega_2$) plane, for the chiral Higgs case, with $m_{\tilde{t}_1}=500$ GeV [left panel] and $m_{\tilde{t}_1}$=1 TeV [right panel], while  $\tan \beta =10$, $A_t=-400$ GeV. The one-loop Higgs mass with tree-level D-terms corrections $m_{h,\,1}$ is plotted.} 
\label{etaversusomega500}
\end{center}
\end{figure}

\noindent The masses of the Higgs states are adjusted accordingly
\small
\begin{align}
m^{2{\rm,\,chir} }_{h^0,H^0}
&=\frac{1}{2}\left(m_A^2 + m_Z^2\right)+\left(C+D\right)\nonumber\\
&\quad\mp\frac{1}{2} \sqrt{\left(m_A^2\! - \! m_Z^2+\frac{2\left(C\!-\! D\right)}{\cos(2\beta)}\right)^2\! \! \! c^2(2\beta)+ \!\left(m_A^2+ m_Z^2 -2m^2_{\Omega}\right)^2\! \! \! s^2(2\beta)}\,,\\
 m_A^{2{\rm,\,chir} }&\equiv \frac{2B_{\mu}}{\sin 2\beta}=m_{H_u}^2+m_{H_d}^2 + 2|\mu|^2+C+D+m^2_{\Omega}\,,\\ 
 m^{2{\rm,\,chir} }_{H^{\pm}}&=m_A^2 +m_W^2(1 - \Omega_2)\,,
 \end{align}\normalsize
where $c^2(2\beta)=\cos^2(2\beta)$, $s^2(2\beta)=\sin^2(2\beta)$. The non-decoupling D-terms in this model leads to a shift to the tree level mass, that in the leading order in the $1/\tan\beta$ expansion is given by
\begin{equation}
    m^2_{h,0}\simeq \left[m_Z^2+\left(\frac{\frac{3}{5}g^2_1\xi_1^2\Omega_1+g^2_2\xi_2^2\Omega_2}{4}\right)v^2\right] + \mathcal{O}(\frac{1}{\tan^2 \beta},\xi_i)\, .
\label{eq:mh1QEW2}    
      \end{equation}
In the following, for simplicity we take $\Omega\equiv\Omega_1=\Omega_2$ and $\xi\equiv\xi_1=\xi_2$.
In fig. \ref{ChiralXiMstop1} it is plotted the Higgs mass from eq. \eqref{Eq:1loopmhMSSM} with the tree-level D-term corrections from \eqref{eq:mh1QEW2} in the ($m_{\tilde{t}_1}$, $X_t$) plane for different values of $\xi$ or $\Omega$. Also in the case the $125.5$ GeV contour lines show that the D-term contribution lower the minimal stop masses required for a given value of $X_t$.  In fig. \ref{etaversusomega500}, in a fashion similar to \cite{Craig:2012bs}\footnote{Note the different notation.}, we show the $m_{h}$ contour lines in the ($\xi$, $\Omega$)-plane for $m_{\tilde{t}_1}=500$ GeV and 1 TeV.

%
In the chiral Higgs case, too, the explicit supersymmetry breaking in the low energy effective theory leads to a reasoning similar to the one discussed in section \ref{FTdterms}.


\section{Higgs couplings determination at LHC and ILC}\label{section:HiggsatILC}
Non-decoupling D-terms induced by the quiver extensions of the MSSM, a part of shifting the tree level masses of the scalars of the theory (see appendix \ref{app:generalDterms}), have direct impact also on several physical quantities as, for instance, the $h\rightarrow\gamma\gamma$ decay branching ratio \cite{Huo:2012tw} or the Higgs boson couplings to fermions and gauge bosons \cite{Arvanitaki:2011ck,Blum:2012ii,Blum:2012kn,Craig:2012bs}. We will study the latter effects, 
estimating the dependence of the deviations from the SM couplings on the additional D-terms, in the light of the precise determination of Higgs boson couplings at current and future colliders. Let us then first define the ratio of the Higgs (the lightest eigenstate $h$) coupling normalised by that of the Standard Model couplings:
\begin{equation}
 \kappa_U=g_U/g_U^{SM}\,,\quad \kappa_D=g_D/g_D^{SM}\,,\quad \kappa_V=g_V/g_V^{SM}\,,
\end{equation}
for any up(down)-type fermion $U=u,c,t$ ($D=d,s,b,e,\mu,\tau$), or gauge boson $V=W^{\pm},Z$.

A standard way to express these ratios, or scaling factors, in a 2HDM models of type-II such as the MSSM, is to write them in terms of the angles $\beta$ and $\alpha$,\begin{equation}
 \kappa_{D}\equiv -\frac{\sin \alpha}{\cos \beta} \ \ , \ \  \kappa_{U}\equiv\frac{\cos \alpha}{\sin \beta} \ \ , \ \  \kappa_V\equiv\sin (\beta-\alpha)\,, \label{RatiosAlphaBeta}
\end{equation} where $\alpha$  is defined as the mixing angle of the Higgs mass eigenstates,
\begin{equation}
 \binom{h_0}{H_0}=\sqrt{2}\binom{-\sin \alpha  \ \ \ \cos \alpha }{ \ \ \cos \alpha  \ \ \ \sin \alpha }\binom{\text{Re} \ H^0_d \ }{\text{Re} \ H^0_u \ }\,.
\end{equation}

The SM is recovered for $\sin\alpha=-\cos\beta,\,\cos\alpha=+\sin\beta$. We can express $\kappa_t,\,\kappa_V$ in terms of $\tan\beta$ and $\kappa_b$ (not considering wrong mixings $\Delta_b$ coming from loop effects) \cite{Blum:2012ii}:
\be \label{why}
 \kappa_t=\sqrt{1-\frac{\kappa_b^2-1}{\tan^2\beta}}\,,  \quad \kappa_V=\frac{\tan \beta}{1+\tan^2 \beta}\left(\frac{\kappa_b}{\tan \beta}+\sqrt{1+\tan^2\beta-\kappa_b^2}\right)\,. 
\ee
The relations \eqref{RatiosAlphaBeta} are exact, however a more transparent general expression for the scaling factors can be obtained looking at the specific model considered. 
We study models in the decoupling limit for large $\tan\beta$. A procedure to rewrite the Higgs couplings in this regime is to start from the general 2HDM Higgs scalar potential, equation \eqref{2HDMPotential} and integrate out the heavier states identified to the Higgs doublet $H_d$, see also \cite{Blum:2012ii,Blum:2012kn,Craig:2012bs}. The Higgs couplings can be read from the effective Lagrangian after having integrated out $H^0_d$ and, after a perturbative expansion in $1/\tan\beta$, $\kappa_b=\kappa_{\tau}$ are
\begin{equation}
 \kappa_b\simeq\left(1-\frac{m_h^2}{m_H^2}\right)^{-1}\left( 1-\frac{\left[\lambda_3+\lambda_5\right] v^2}{m_H^2-m_h^2}\right)+\ldots\,, \label{eq:cbgeneral}
\end{equation}
where we adopt the definitions from \refe{2HDMPotential}  ($v^2\sim(246 \  \text{GeV})^2$) and the ellipsis denote nonzero $\lambda_7$ contributions from F-term like enhancements that are null in the MSSM, $\mathcal{O}(1/\tan^2\beta)$ corrections and possible ``wrong sign" couplings coming from 1-loop contributions. Finding the right $\kappa_b$ expressions for our quiver models is straightforward, substituting into \eqref{eq:cbgeneral} the corresponding $\lambda_3,\,\lambda_5$. For the vector Higgs case $\lambda_3,\,\lambda_5$ are obtained by the MSSM relations \eqref{MSSMPotentialparameters} with the additional contributions \eqref{eq:D-terms}, giving $\lambda_3+\lambda_5=[g^2_2(1+\Delta_2) + \frac{3}{5}g^2_1(1+\Delta_1)]/4$, such that
\begin{equation}
 \kappa^{\text{vector}}_b\simeq \left( 1- \frac{m_h^2}{m_H^2}\right)^{-1}\left(1+ \frac{[g^2_2(1+\Delta_2) + \frac{3}{5}g^2_1(1+\Delta_1)]v^2}{4\left(m_H^2-m_h^2\right)} \right)\,.\label{CbVectorCase}
\end{equation}
For the chiral Higgs case, using instead the additional contributions \eqref{eq:D-termsC}, one obtains $\lambda_3+\lambda_5=[g^2_2(1 -\Omega_2) + \frac{3}{5}g^2_1(1-\Omega_1)]/4$, and
\begin{equation}
 \kappa^{\text{chiral}}_b\simeq \left( 1- \frac{m_h^2}{m_H^2}\right)^{-1}\left(1+ \frac{[g^2_2(1 -\Omega_2) + \frac{3}{5}g^2_1(1-\Omega_1)]v^2 }{4\left(m_H^2-m_h^2\right)} \right)\,.\label{CbChiralCase}
\end{equation}
In these two cases, the MSSM limit can be obtained by setting the non-decoupling D-term contributions to zero, respectively $\Delta_i=0$ and $\Omega_i=0$. 
\begin{figure}[htb]\centering
\subfigure[]{\label{LHCvsILCSensVec}\includegraphics[width=.49\textwidth]{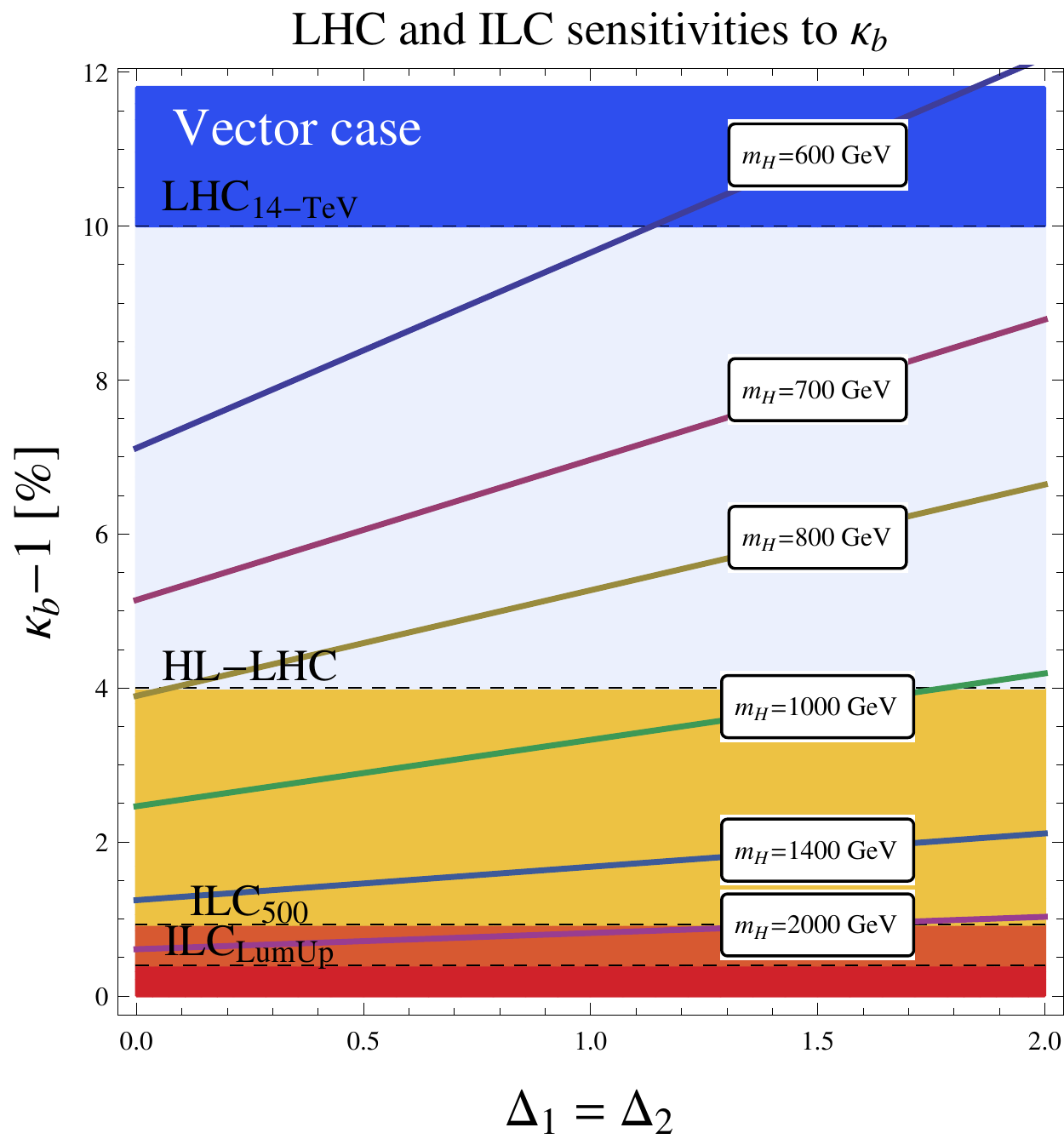}}
\subfigure[]{\label{ILCSensVec}\includegraphics[width=.49\textwidth]{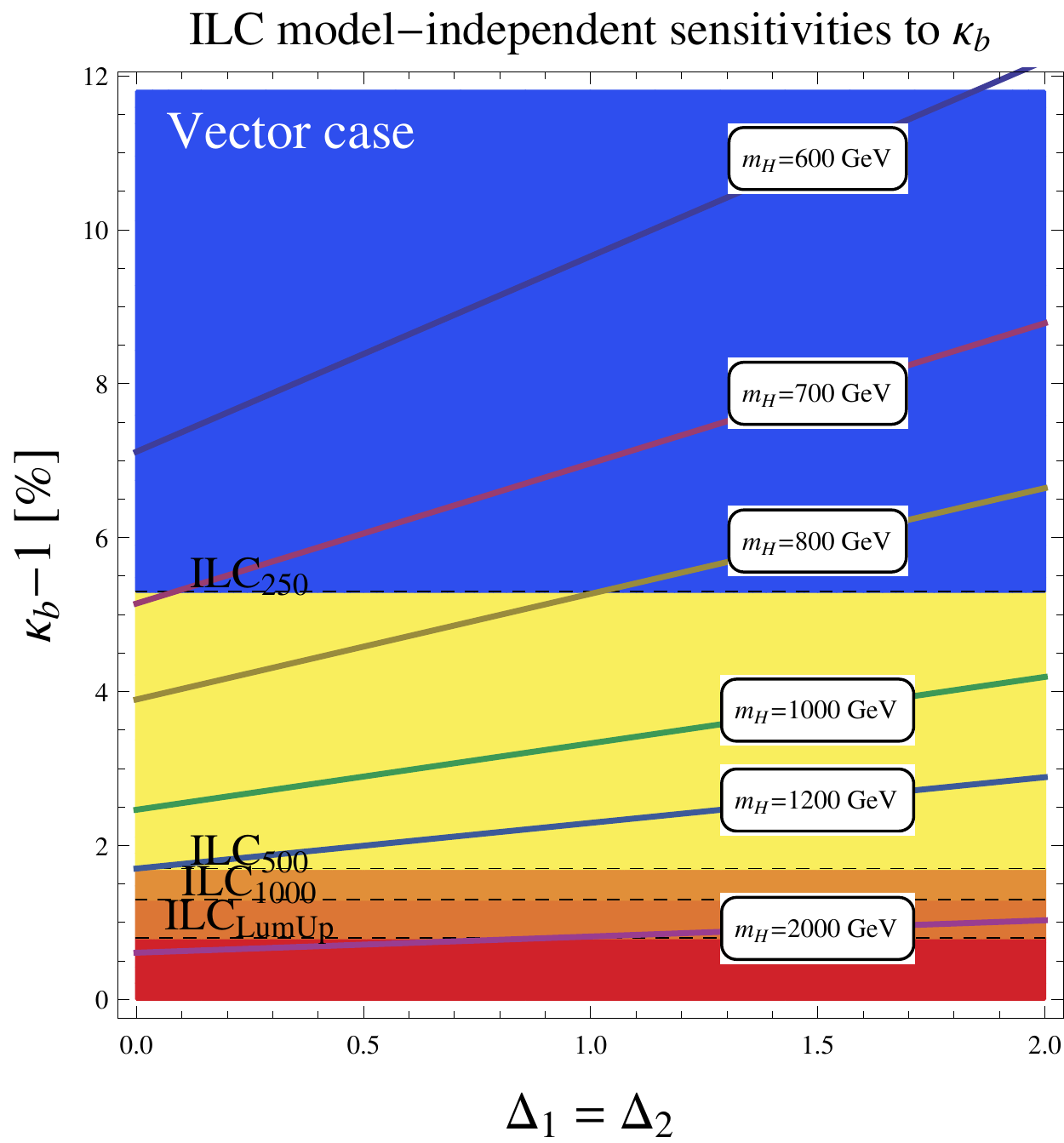}}
\caption{Vector case:  relative enhancements $\kappa_b-1$ of the Higgs bottom couplings for the chiral-Higgs case with respect to the SM are displayed in solid lines, in [\%] as function of $\Delta_1=\Delta_2$, for different values of $m_H$ [GeV]. (a) In dashed lines, the contours of the expected accuracies on the scaling factors $\kappa_b$ at at the LHC, HL-LHC and ILC, from \cite{Dawson:2013bba} and table \ref{HigCoupLHCILC}, centered on the SM value 0. The accuracies assume  no non-SM production and decay modes and assumes universality ($\kappa_u\equiv \kappa_t= \kappa_c$, $\kappa_d\equiv \kappa_b=\kappa_s$ and $\kappa_l\equiv \kappa_{\tau}=\kappa_{\mu}$). (b)   In dashed lines, the contours of the model-independent ILC sensitivities for each run from \cite{Asner:2013psa}, see table \ref{HigCoupILC}, centered on the SM value 0.
\label{ILCSensVec_both}}
\end{figure}

\begin{figure}[htb]\centering
\subfigure[]{\label{EllipseVec}\raisebox{+0.08\height}{\includegraphics[width=.49\textwidth]{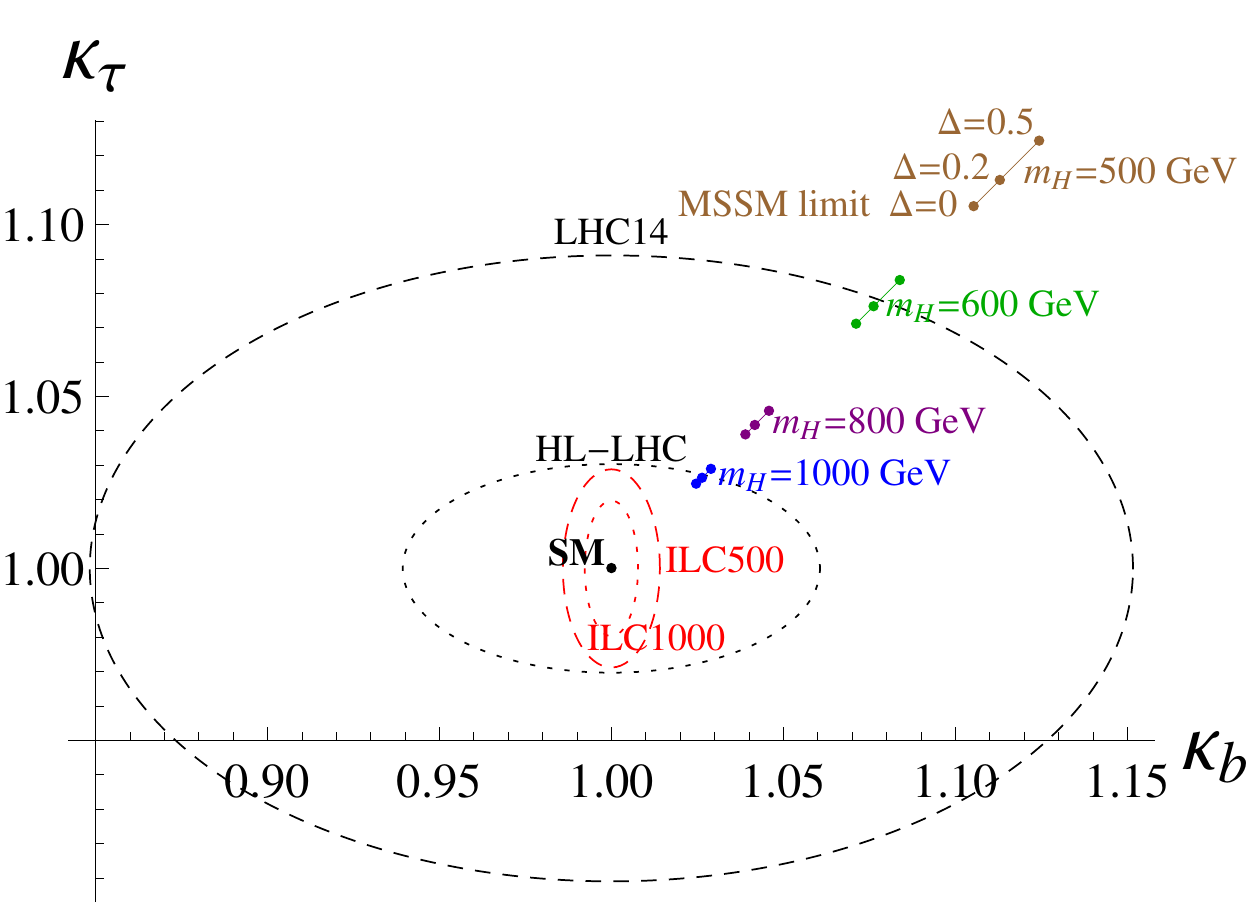}}}
\subfigure[]{\label{Chi2CouplingsVector}\includegraphics[width=.49\textwidth]{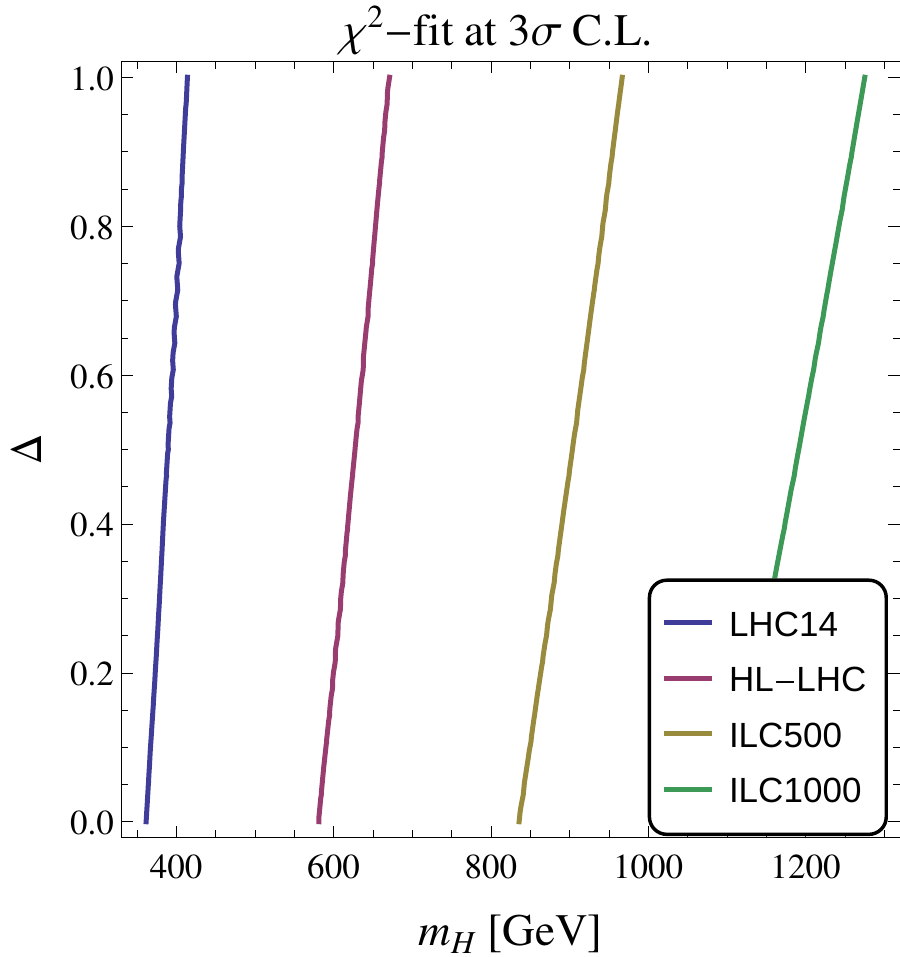}}
\caption{Vector-Higgs case: experimental sensitivity to coupling deviations from the Standard Model, assuming no correlation between $\kappa_i$ measures. (a) $(\kappa_b,\,\kappa_{\tau})$ for $\Delta$=0 (SM-limit), 0.1, 0.2, 0.5, at different values of $m_H=500,\,600,\,800,\,1000,\,1200$ GeV. The experimental sensitivity, centered in the SM value $(\kappa_b,\,\kappa_{\tau})$=1, is represented by 1$\sigma$-confidence ellipses: black dashed for LHC at 14 TeV and 300 fb$^{-1}$, black dotted for HL-LHC at 3000 fb$^{-1}$ at 14 TeV, red dashed ILC at 500 GeV and red dotted for ILC at 1000 GeV. (b) $\chi^2$-test of $\kappa_W,\,\kappa_Z,\,\kappa_{\tau},\,\kappa_b,\,\kappa_t$ in the $(m_H,\,\Delta)$-plane at the different experiments: areas on the left of the solid lines are not consistent with the SM at 3$\sigma$-confidence level.\label{ILCSensVec_FIT}}
\end{figure}

\begin{figure}[htb]\centering
\subfigure[]{\label{LHCvsILCSensChir}\includegraphics[width=.49\textwidth]{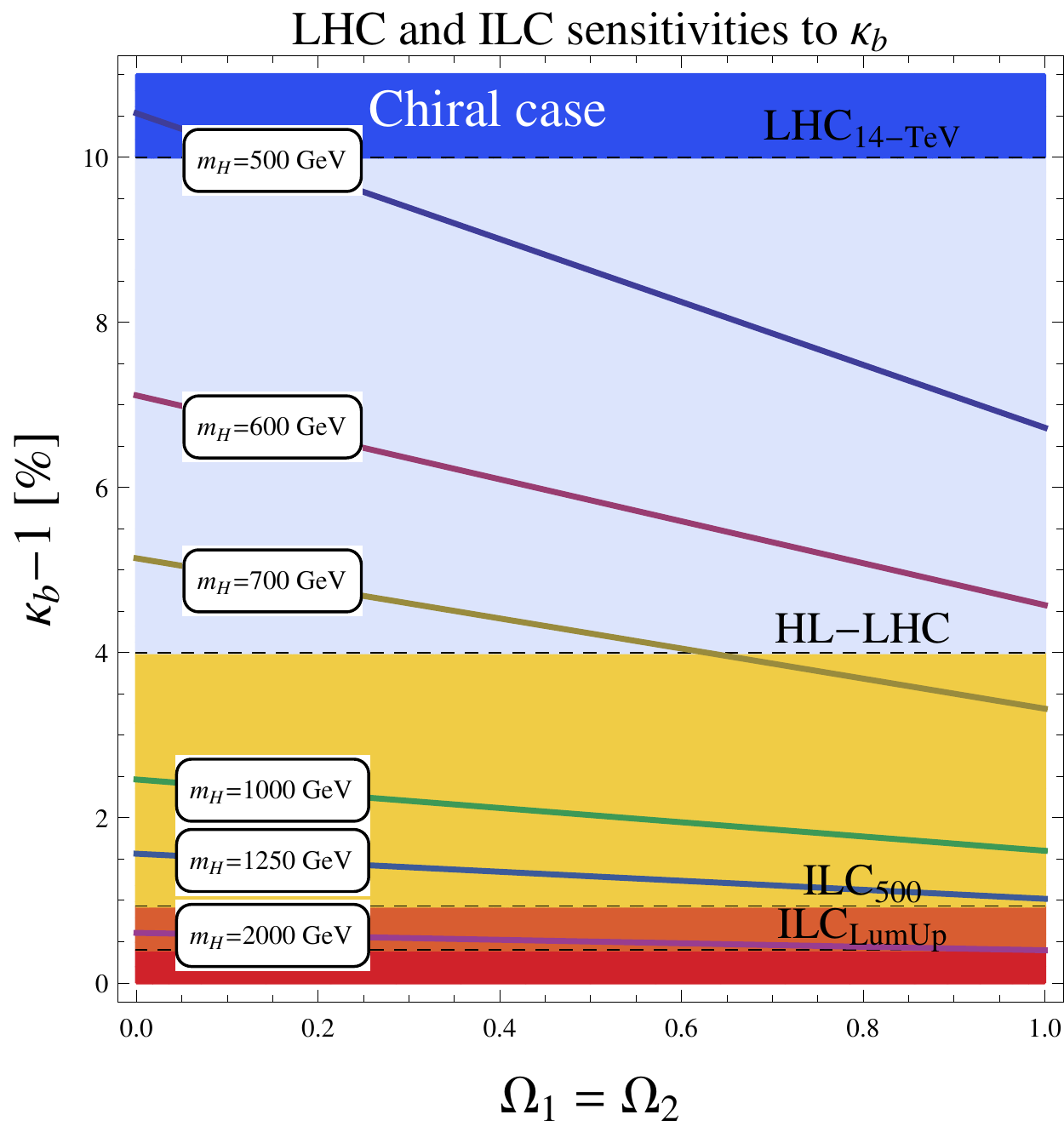}}
\subfigure[]{\label{ILCSensChir}\includegraphics[width=.49\textwidth]{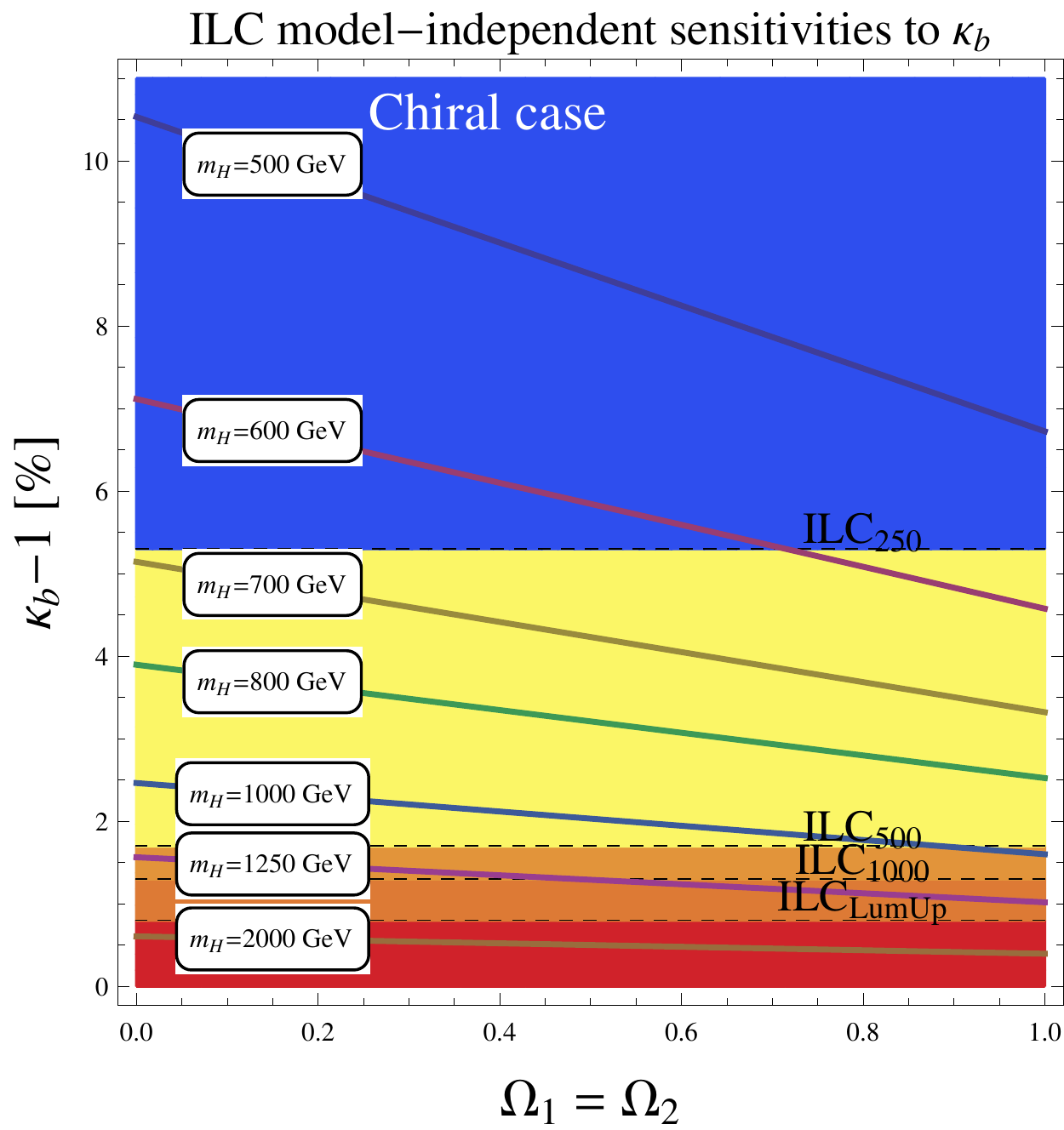}}
\caption{Chiral-Higgs case: relative enhancements $\kappa_b-1$ of the Higgs bottom couplings for the chiral-Higgs case with respect to the SM are displayed in solid lines, in [\%] as  a  function of $\Omega_1=\Omega_2$ for different values of $m_H$ [GeV].  (a) In dashed lines, the contours of the expected accuracies on the scaling factors $\kappa_b$ at the LHC, HL-LHC and ILC, from \cite{Dawson:2013bba} and table \ref{HigCoupLHCILC}, centered on the SM value 0. The accuracies assume  no non-SM production and decay modes and assumes universality ($\kappa_u\equiv \kappa_t= \kappa_c$, $\kappa_d\equiv \kappa_b=\kappa_s$ and $\kappa_l\equiv \kappa_{\tau}=\kappa_{\mu}$). Correlations are neglected. (b) In dashed lines, the contours of the model-independent ILC sensitivities for each run from \cite{Asner:2013psa}, see table \ref{HigCoupILC}, centered on the SM value 0.\label{ILCSensChir_both}}
\end{figure}

\begin{figure}[thb!]\centering
\subfigure[]{\label{EllipseChir}\raisebox{+0.08\height}{\includegraphics[width=.49\textwidth]{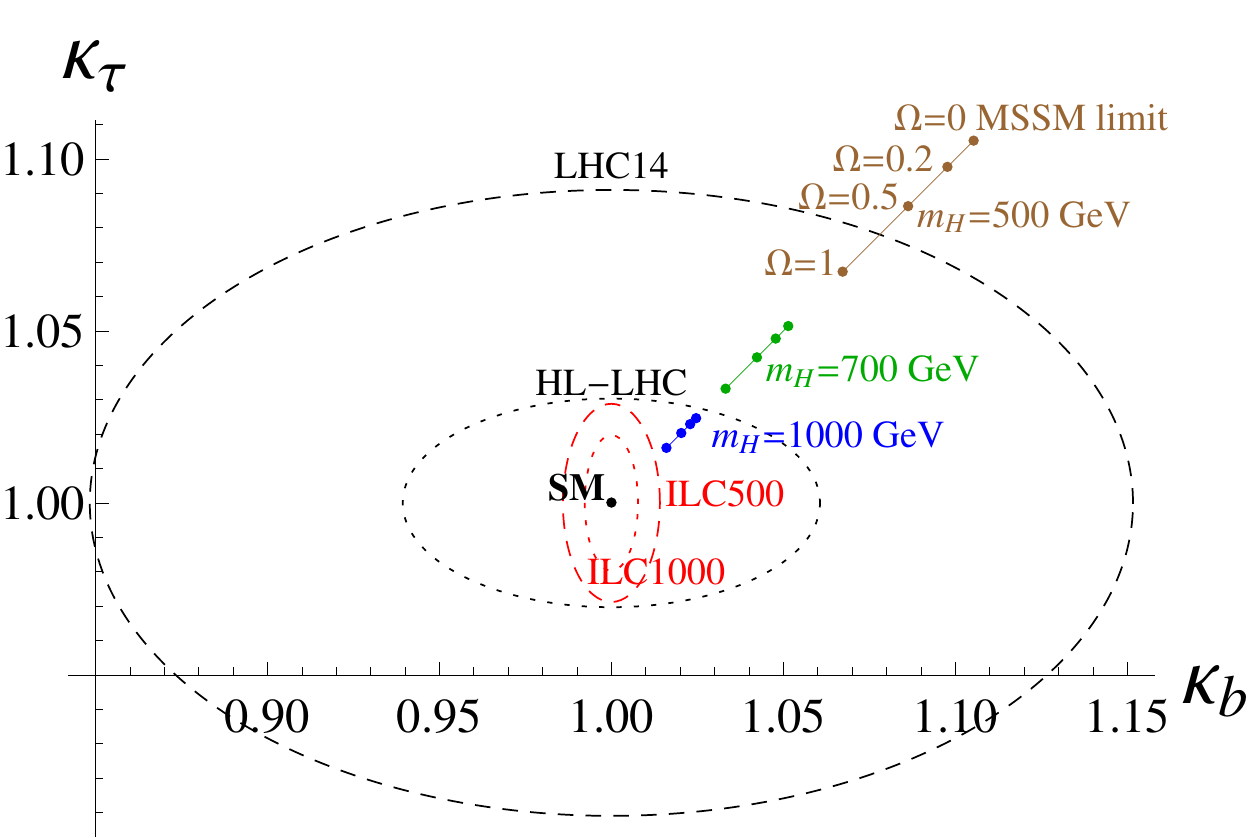}}}
\subfigure[]{\label{Chi2CouplingsChiral}\includegraphics[width=.49\textwidth]{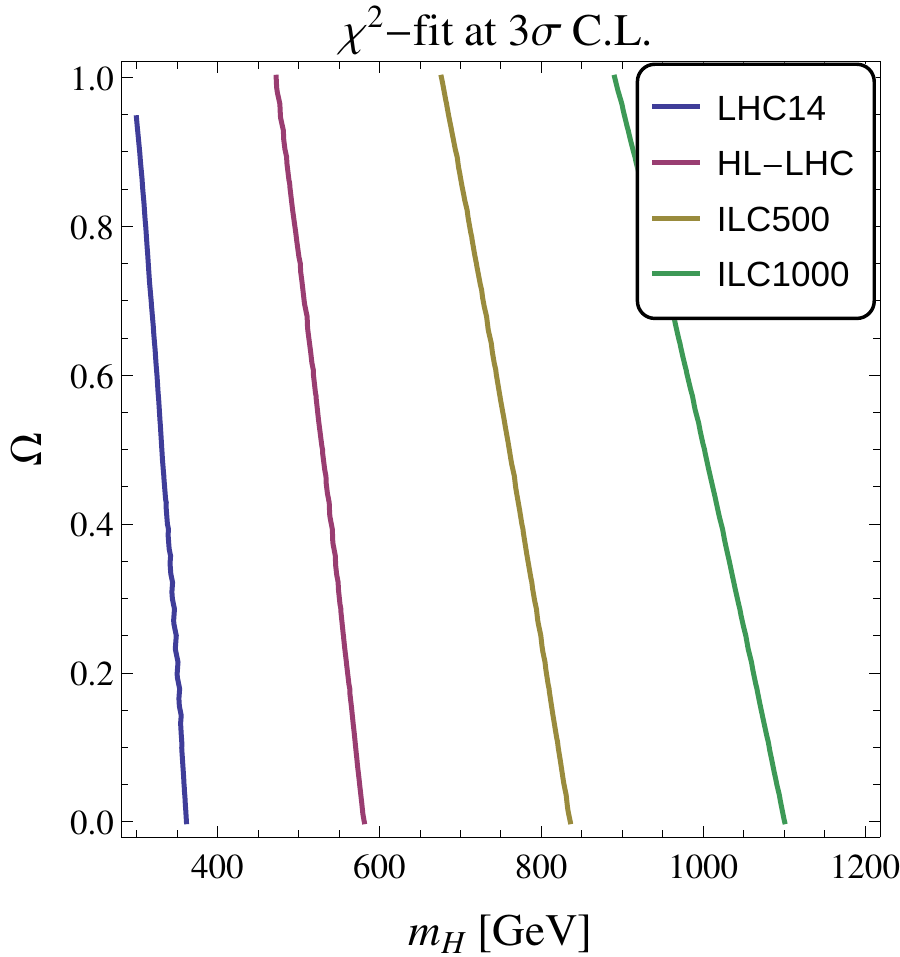}}
\caption{Chiral-Higgs case: experimental sensitivity to coupling deviations from the Standard Model, assuming no correlation between $\kappa_i$ measures. (a) $(\kappa_b,\,\kappa_{\tau})$ for $\Omega$=0 (SM-limit), 0.2, 0.5, 1 at different values of $m_H=500,\,600,\,800,\,1000,\,1200$ GeV. The experimental sensitivity, centered in the SM value $(\kappa_b,\,\kappa_{\tau})$=1, is represented by 1$\sigma$-confidence ellipses: black dashed for LHC at 14 TeV and 300 fb$^{-1}$, black dotted for HL-LHC at 3000 fb$^{-1}$ at 14 TeV, red dashed ILC at 500 GeV and red dotted for ILC at 1000 GeV. (b) $\chi^2$-test of $\kappa_W,\,\kappa_Z,\,\kappa_{\tau},\,\kappa_b,\,\kappa_t$ in the $(m_H,\,\Delta)$-plane at the different experiments: areas on the left of the solid lines are not consistent with the SM at 3$\sigma$-confidence level.\label{ILCSensChir_FIT}}
\end{figure}

                
It is important to understand how these D-term enhanced deviations from the SM couplings could be detected, as a signature for the considered quiver-models at present and future colliders, see figures \ref{ILCSensVec_both}, \ref{ILCSensVec_FIT}, \ref{ILCSensChir_both}, \ref{ILCSensChir_FIT}.
At the LHC only ratios between different Higgs couplings can be determined, therefore coupling determination is possible only in the framework of a specific model. For example, taking some minimal assumptions on the underlying model, as explained in \cite{Dawson:2013bba}, one can obtain $\kappa_b$ from a constrained 7-parameter fit assuming no non-SM production and decay modes and assuming generation universality ($\kappa_u\equiv \kappa_t= \kappa_c$, $\kappa_d\equiv \kappa_b=\kappa_s$ and $\kappa_l\equiv \kappa_{\tau}=\kappa_{\mu}$). This is listed in table \ref{HigCoupLHCILC}, where the coupling determination uncertainties at LHC at 14 TeV ($\int\mathcal{L}dt=300\text{ fb}^{-1}$) and High Luminosity 	LHC (HL-LHC, $\int\mathcal{L}dt=3000\text{ fb}^{-1}$) are compared with some expectations at the International Linear Collider (ILC). 
\begin{table}[ht!]   
\begin{center}\footnotesize
\begin{tabular}{|c|cccccc|}
\hline
 \hline 
 & \bf LHC 14&\bf HL-LHC& \bf $\mbox{ILC}_{500}$& \bf $\mbox{ILC}_{\rm500-LumUp}$& \bf $\mbox{ILC}_{1000}$&  \bf $\mbox{ILC}_{\rm1000-LumUp}$\! \\\hline
 $\! \! \kappa_W\! \! $&4\,--6 \% & \!2\,--5 \%&0.39 \%&0.21 \%&0.21 \%&0.2 \%\!\!  \\\hline 
  $\! \! \kappa_Z\! \! $&4\,--6 \% & \!2\,--4 \%&0.49 \%&0.24 \%&0.5 \%&0.3 \%\!\!  \\\hline 
   $\! \! \kappa_l=\kappa_{\tau}\! \! $&6\,--8 \% & \!\,2\,--5 \%&1.9 \%&0.98 \%&1.3 \%&0.72 \%\!\!  \\\hline 
    $\! \! \kappa_d=\kappa_b\! \! $&10\,--13 \% & \!4\,--7 \%&0.93 \%&0.60 \%&0.51 \%&0.4 \%\!\!  \\\hline
        $\! \! \kappa_u=\kappa_t\! \! $&14\,--15 \% & \!7\,--10 \%&2.5 \%&1.3 \%&1.3 \%&0.9 \%\!\!  \\\hline \hline 
\end{tabular} \end{center}\caption{Expected precisions on $\kappa_b$ at $1\sigma$, in \%, from a constrained 7-parameter fit assuming no non-SM production and decay modes and assuming universality ($\kappa_u\equiv \kappa_t= \kappa_c$, $\kappa_d\equiv \kappa_b=\kappa_s$ and $\kappa_l\equiv \kappa_{\tau}=\kappa_{\mu}$), as reported in \cite{Dawson:2013bba}. LHC corresponds to 300 fb$^{-1}$ at 14 TeV, HL-LHC at 3000 fb$^{-1}$ at 14 TeV.   }\label{HigCoupLHCILC} 
\end{table}
On the other hand, at future $e^+e^-$-colliders as the ILC, the Higgs \emph{total} width and the Higgs couplings 
can be determined in a model-independent way. This is possible by exploiting the recoil methods that allow for a decay independent determination of the Higgsstrahlung process production $e^+e^-\rightarrow HZ$, a quantity that enters many observables \cite{Asner:2013psa}.  With respect to estimates with minimal model assumption, there are slightly higher uncertainties.
This is reported in table \ref{HigCoupILC}, where we show the estimated ILC accuracies on the Higgs couplings, assuming the theoretical uncertainties to be equal to 0.5\% for the ILC stages at $\sqrt{s}=$250, 500, 1000 GeV and for the luminosity upgrade $\mbox{ILC}_{\rm LumUp}$ at 250, 500, 1000 GeV, from \cite{Asner:2013psa}, that may be further improved \cite{Peskin:2013xra}. 
Since ILC measurements are dominated by statistical errors, they are improved with increasing statistics, in contrast with Higgs determinations in the High-Luminosity LHC that are dominated by systematic errors.


\begin{table}[ht!]
\begin{center}\footnotesize
\begin{tabular}{|c|cccc|}
\hline
 \hline 
 &\bf $\mbox{ILC}_{250}$&\bf$\mbox{ILC}_{500}$&\bf$\mbox{ILC}_{1000}$&\bf $\mbox{ILC}_{\rm LumUp}$\\\hline
 $\kappa_W$&4.9 \%&1.2 \%&1.1 \%&0.6 \%\\
 $\kappa_Z$&1.3 \%&1.0 \%&1.0 \%&0.5 \%\\
 $\kappa_{\tau}$&5.8 \%&2.4 \%&1.8 \%&1.0 \%\\
 $\kappa_b$&5.3 \%&1.7 \%&1.3 \%&0.8 \%\\
 $\kappa_t$&--&14 \%&3.2 \%&2.0 \%\\
\hline \hline
\end{tabular} \end{center}\caption{Expected accuracies on the coupling scaling factors $\kappa_i$ at $1\sigma$, in \%, for a completely model-independent fit assuming theory errors $\Delta F_i/F_i=0.5\%$, from the ILC Higgs White Paper \cite{Asner:2013psa}.}\label{HigCoupILC} 
\end{table}
In figure \ref{ILCSensVec_both} we plot how the LHC and ILC may detect deviations from the SM Higgs bottom coupling due to non-decoupling D-terms in a vector Higgs quiver extension of the MSSM. 
The relative enhancement with respect to the SM Higgs bottom coupling, $\kappa_b-1$, is plotted as a function of $\Delta$ for different values of the heavier neutral CP-even Higgs mass $m_H$ (see eq. \eqref{CbVectorCase}). The non-decoupling D-terms in the vector Higgs case enhance the deviation from the SM with respect to the MSSM limit $\Delta=0$, while larger values for $m_H$ clearly suppress these effects. Furthermore, in figure \ref{ILCSensVec_both} a value of $\kappa_b-1$ that lies above a contour line corresponds to a deviation from the SM that can be detected, once the Higgs bottom coupling at the corresponding machine run is measured.
In figure \ref{LHCvsILCSensVec}, the horizontal dashed contour lines correspond to the LHC and ILC 1$\sigma$-confidence level sensitivities for $\kappa_b$ determination with the minimal model assumptions in table \ref{HigCoupLHCILC}, centered on the SM value $\kappa_b^{SM}-1=0$. In \ref{ILCSensVec}, the sensitivities for ILC model-independent $\kappa_b$ determination are displayed. At the LHC at 14 TeV, deviations triggered by $\Delta\sim\mathcal{O}(1\text{-}2)$ may be detected for a  $m_H\lesssim600$ GeV, while for $m_H\leq1$ TeV, these deviations may be detected at the HL-LHC.  
Coupling enhancements due to $\Delta\sim\mathcal{O}(0.1\text{-}0.2)$, more suitable according to the top-down approach in \cite{Bharucha:2013ela}, are (just) discernible at the  HL-LHC for $m_H\lesssim800$ GeV.
The ILC running at 500 GeV may be sensitive to these ranges of $\Delta$ for $m_H \sim 1$ TeV, while for the high luminosity configuration at 1000 GeV ($\int\mathcal{L}dt=\mathcal{O}(5000)\text{ fb}^{-1}$), this is valid up to $m_H\sim2$ TeV, showing the power of this experiment in the study of the Higgs scalar potential.

The deviation from the SM of $\kappa_b$ in figure \ref{ILCSensVec_both} alone cannot be used for claiming a BSM underlying model, as it can merely be due to statistical effects. Therefore, in figure \ref{EllipseVec} we show the non-decoupling D-terms triggered deviations in $\kappa_b$ and $\kappa_{\tau}$: the points lying outside the $1\sigma$-confidence ellipses for each experiment is displayed.\footnote{A similar kind of analysis, for general 2HDM models, may be found in \cite{Kanemura:2014bqa}.} In figure \ref{Chi2CouplingsVector} we perform a $\chi^2$-fit to the SM values of $\kappa_W,\,\kappa_Z,\,\kappa_{\tau},\,\kappa_b,\,\kappa_t$ in the $(m_H,\,\Delta)$-plane:  the areas on the left of the solid lines are not consistent with the SM at 3$\sigma$-confidence level.  
As deviations from the SM value 1 for $\kappa_W,\,\kappa_Z,\,\kappa_t$ are relatively mild in these models, see eq. \eqref{why}, 
in particular considering the achievable accuracy in these quantities, the main contribution to the $\chi^2$ result comes from $\kappa_b$ and $\kappa_{\tau}$, as they present large deviations and a relatively good resolution.
One can see that at the first run of the LHC deviations from the Standard Model only for a relatively light $H$, with mass up to $m_H\simeq350$-$400$ GeV are observable and the luminosity upgrade is needed to explore the parameter space up to decoupling masses $m_H\lesssim500$ GeV for any value of $\Delta$. At the ILC, instead, deviations from the SM for $m_H$ up to 700 (900) GeV  at $\sqrt{s}=$500 (1000) GeV.
In both plots in fig. \ref{ILCSensVec_FIT} we do not take into account any experimental correlations between the determinations of $\kappa_i$.


In the chiral Higgs quiver case, the D-terms triggered deviations of $\kappa_b$ (in particular), have an opposite behaviour compared to the vector case. Here, the D-term contributions are negative, see eq. \eqref{CbChiralCase}, pushing the Higgs couplings closer to the SM value. Therefore for increasing $\Omega$, the deviations of the couplings from the SM get less detectable with respect to the MSSM limit (for $\Omega=0$), see figures \ref{ILCSensChir_both} and \ref{EllipseChir}. Figure \ref{Chi2CouplingsChiral} shows how, for $\Omega\sim\mathcal{O}(1)$, the sensibility to the deviation of couplings is reduced at the LHC by $\sim$ 50 GeV and by $\sim$ 100 GeV at the ILC. 

Once a deviation in the couplings from the SM is detected, one should address the question about which (BSM) supersymmetric model has been observed. For this, the measurement of the couplings alone is not sufficient, but also the detection of $H$ and the measurement of its mass $m_H$ are fundamental, as one can decouple the $\Delta$ or $\Omega$ measurement, see equation \eqref{CbVectorCase}.

\section{Discussion and Conclusions}\label{section:conclusion}
In the Minimal Supersymmetric Standard Model (MSSM) a 125.5 GeV Higgs implies large radiative corrections from stops or large stop mixing. 
Such a requirement can be relaxed in the framework of non-decoupling D-terms extensions, in which additional contributions to Higgs quartic couplings enhance the tree-level Higgs mass.
We studied two examples of quiver models that result in two different low energy D-terms extensions of the MSSM: the ``vector Higgs'' case, with both Higgs doublets in the same gauge site, and the ``chiral Higgs'' case, with the Higgs doublets in two different sites.

In the vector Higgs case we concentrated on the region in which the D-term size parameter $\Delta$ is $\sim\mathcal{O}(0.1)$ as it may be preferred in the light of perturbative unification and from a top down approach \cite{Bharucha:2013ela}. For example, for $X_t=0$ GeV, the Higgs mass $m_h=125.5$ GeV is recovered with $m_{\tilde{t}_1}\sim\mathcal{O}(1)$ TeV for $\Delta=0.3$, while for the MSSM limit $\Delta=0$, a $m_{\tilde{t}_1}\sim\mathcal{O}(4)$ TeV is required, showing how non-decoupling D-terms may increase the tree-level mass. Non-decoupling D-terms also modify the couplings of the Higgs boson to fermions and vector bosons with respect to the SM and the MSSM. The measurement of the quantities $k_i=g_i/g_{i, {\rm SM}}$, especially $\kappa_b,\,\kappa_{\tau}$ at the LHC and ILC may be used to discriminate from the SM and the MSSM itself. In the vector Higgs case in particular, considering the decoupling limit $m_h\ll m_H$ and large $\tan\beta$, the coupling ratios $\kappa_b,\,\kappa_{\
tau}$ sensibly increase with respect to the SM and MSSM  for increasing $\Delta$. At the LHC at 14 TeV deviations from the SM may be determined for any value of $\Delta$ only for relatively light $H$, $m_H\leq350$ GeV, as the lighter $H$ the larger is the correction. At the HL-LHC deviations from the SM can be determined for $H$ roughly $200$ GeV heavier. At the ILC at 500 GeV instead deviations from the SM are seen with $m_H\leq800$ GeV for $\Delta=0$ and with $m_H$ up to 900 GeV for $\Delta=0.5$; the improved resolution at the ILC at 1 TeV may push the detectable deviation heavier by another $\sim250$ GeV.
In the Chiral Higgs case the tree-level Higgs mass is enhanced (similarly to the vector Higgs case) with increasing $\xi^2\cdot\Omega$. However $\kappa_b,\,\kappa_{\tau}$ decrease for smaller $\Omega$ and get closer to the SM, the determination of a deviation from the SM becomes more challenging. In particular at the ILC at 500 (1000) TeV for $\Omega=1$ (the maximal value), deviations from the SM can be found only for $m_H\leq650$ (900) GeV, roughly 200 GeV lighter than in the MSSM limit $\Omega=0$.

On the other hand, once deviations from SM couplings are established, further studies are needed to determine the underlying model. For instance, within supersymmetry, in order to distinguish the model from the MSSM, also a precise measurement of $m_H$ is required for obtaining $\Delta$ or $\Omega$. Furthermore, combining these results with electroweak precision measurements where the effects of gauge extensions could be observed, may possibly identify these models. 
In order to be sensitive to a vast range of gauge extended models, we have shown that the precise and largely model-independent measurements of the Higgs couplings at the linear collider is needed.


\acknowledgments  The authors would like to thank K.~Blum and R.T.~D'Agnolo for comments regarding \cite{Blum:2012kn}. M.M. would like to thank A.~Bharucha and A.~Goudelis. S.P. would like to thank P.~Drechsel, K.~Fujii, S.~Kanemura, F.~Moortgat, G.~Nardini, M.~Tonini, M.~de Vries and G.~Weiglein and acknowledges the support of DFG through the grant SFB 676 ``Particles, Strings, and the Early Universe". This work was partially supported by the Foundation for Polish Science International PhD Projects Programme co-financed by the EU European Regional Development Fund. This work has been partially supported by National Science Centre under research grant DEC-2012/04/A/ST2/00099. 


\appendix

\section{General derivation of non-decoupling D-terms }\label{app:generalDterms}

Here we give a derivation of the non-decoupling D-terms and show how they may arise within a two-site quiver model, involving scalars charged under the final symmetry: squarks, sleptons as well as Higgs bosons. We consider the product of two identical (non-)abelian gauge groups $G_A\times G_B$ that breaks to the diagonal subgroup, $G_D$. The canonical kinetic terms for $i$ chiral superfields $A_i$, charged under only $G_A$, and of $j$ chiral superfields $B_j$, charged under only $G_{B}$, are given by:
\begin{equation}
\mathcal{L}\supset \int d^4\theta \Big( \sum_i A^{\dagger}_i e^{g_aV_a}A_i+ \sum_j B^{\dagger}_j e^{g_bV_b}B_j \Big),
 \end{equation}
where $g_a$ and $g_b$, respectively, are the gauge couplings for site $A$ and $B$ and $V_a,\,V_b$ are the corresponding vector multiplets.

After the diagonal breaking $G_A\times G_B\rightarrow G_D$, $V_a$ and $V_b$ recombine into a massless vector multiplet, $V_D$, and a heavy one, $V_H$, that can be written as
\begin{equation}
 V_D=\frac{g_aV_b+g_bV_a}{\sqrt{g^2_a+g^2_b}}   \ \  , \  \   V_H=\frac{-g_aV_a+g_bV_b}{\sqrt{g^2_a+g^2_b}}   \ \ \ .
\end{equation}

$V_H$ obtains a mass through the supersymmetric Higgs mechanism by eating a (complex) chiral superfield $\Phi$, in our case a combination of the linking fields between the sites $A$ and $B$,
\begin{equation}
 \Phi=(t+is) +\sqrt{2}\theta \chi+\theta^2 F_{\Phi}\,.
\end{equation}
The real scalar field $t$ is eaten to give the third degree of freedom to the gauge fields $A_{\mu}$, $s$ remains uneaten, while the Weyl fermion $\chi$ couples to the gaugino $\lambda$ to make a supersymmetric Dirac mass. 
In the K\"ahler potential the corresponding mass term $m_V^2$ for $V_H$ is given by
\begin{equation}
 \mathcal{L}\supset \int d^4 \theta \  m_V^2V^2_H +\dots   \,  .
\end{equation}
Furthermore, the following soft mass terms are added,
\begin{equation}
 \mathcal{L}\supset \int d^4 \theta \ (m_{\chi}m_V^2 \theta^2 + \bar{m}_{\chi}m_V^2 \bar{\theta}^2 -\frac{1}{2}m_V^2m_s^2 \theta^4  )V^2_H+\int d^2 \theta m_{\lambda}W^2_{\alpha} +\int d^2 \bar{\theta} \bar{m}_{\lambda}\bar{W}^2_{\dot{\alpha}}\,,
\end{equation}
where the soft masses $m_{\chi},m_s^2,m_{\lambda}$ respectively parameterise the soft breaking of the fermion $\chi$, the real uneaten scalar $s$ and the usual Majorana soft mass for the gaugino $\lambda$. 
Therefore the K\"ahler potential may be written to leading order in $V_H$ as
\begin{equation}
 K_{H}\supset g_d\left(\frac{g_a}{g_b}\right)\mathcal{J}_a V_H +  g_d\left(\frac{g_b}{g_a}\right)\mathcal{J}_b V_H + \dots  \ \    \ .
\end{equation}
$\mathcal{J}_{a/b}$ are the current multiplets, satisfying the constraint $D^2 \mathcal{J}=0$, that contain all the fields charged under site $A$ or site $B$:
\begin{equation}
  {\cal J}^c = J^c + i \theta j^c - i \bar \theta  \bar j^c -
\theta\sigma^\mu\bar \theta j_\mu^c +
\frac{1}{2}\theta\theta\bar\theta\bar \sigma^\mu\partial_\mu j^c -
\frac{1}{2}\bar\theta\bar\theta\theta \sigma^\mu\partial_\mu \bar
j^c - \frac{1}{4}\theta\theta\bar\theta\bar\theta \Box J^c\,,
\end{equation}
with the leading term being the current of scalars $J^c = \sum_i \phi_i^\dag T^c \phi_i$, where $\phi_i$ are the collection of all scalars charged under the gauge group and $c$ is the generator index. The effective lagrangian after integrating out the heavy vector field $V_H$, is then of the form
\begin{equation}
 \mathcal{L}_{\text{eff}}=\int d^4\theta \left( \sum_i A^{\dagger}_i e^{g_DV_D}A_i+ \sum_j B^{\dagger}_j e^{g_DV_D}B_j \right)+\mathcal{O}\,.
\end{equation}
$\mathcal{O}$ is the most general expression for the non-decoupled D-terms,
\begin{equation}
 \mathcal{O}=g_D^2\int d^4 \theta \left(\frac{1}{m_V^2}-\frac{m_s^2\theta^4}{m_V^2 + m_s^2}\right)\sum_A  \Big[\left(\frac{g_a}{g_b}\right)\mathcal{J}^A_a  - \left(\frac{g_b}{g_a}\right)\mathcal{J}^A_b\Big]^2\,, \label{superoperator}
\end{equation}
with a sum over $A$ generators. The associated non-decoupling D-term corresponds then to the $\theta^4$ term in the round brackets of eq. \eqref{superoperator}, while the currents in the square brackets reduce simply to ($\frac{1}{8}$) the scalar current for this $\theta^4$  term.

Passing explicitly to the case of quiver extensions of the MSSM, the diagonal gauge group coupling $g_D$ corresponds to the SM coupling $g_{SM}$ and the symmetry breaking consists in
\begin{equation}
 SU(2)_{A}\times SU(2)_{B}\rightarrow SU(2)_L,\,\,\,\,\,\,\,\,\,\,U(1)_{A}\times U(1)_{B}\rightarrow U(1)_Y\,.
\end{equation}

In the case of a model in which all MSSM fields are on site A, charged under $G_A$ the scalar currents are given by
\begin{align}
 J_{U(1)_A}&=\frac{1}{2} H^{\dagger}_u H_u-\frac{1}{2} H^{\dagger}_dH_d-\frac{1}{2}\tilde{l}^{\dagger}\tilde{l}+\frac{1}{6}\tilde{q}^{\dagger} \tilde{q} + \frac{1}{3} \tilde{d}^{\dagger}\tilde{d} -\frac{2}{3}\tilde{u}^{\dagger}\tilde{u} +\tilde{e}^{\dagger}\tilde{e}\,,\nonumber\\
 J_{U(1)_B}&=0,    \\
 J^A_{SU(2)_A}&=\frac{1}{2}\left( H^{\dagger}_u\sigma^A  H_u+ H^{\dagger}_d \sigma^A H_d +\tilde{q}^{\dagger}\sigma^A \tilde{q} +\tilde{l}^{\dagger}\sigma^A \tilde{l}\right)\,,\nonumber\\ 
 J^A_{SU(2)_B}&=0\,,
\end{align}
with all flavour and colour indices implicitly traced.
For the case of split generations (see for instance \cite{Batra:2004vc}), in which the 3rd generation and $H_u,H_d$ are charged under $G_A$ and the first two generations under $G_B$, one finds
\begin{align}
 J_{U(1)_A}&=\frac{1}{2}H^{\dagger}_uH_u-\frac{1}{2} H^{\dagger}_dH_d + \left[-\frac{1}{2}\tilde{l}^{\dagger}\tilde{l}+\frac{1}{6}\tilde{q}^{\dagger}\tilde{q}+\frac{1}{3}\tilde{d}^{\dagger}\tilde{d}-\frac{2}{3}\tilde{u}^{\dagger}\tilde{u} +\tilde{e}^{\dagger}\tilde{e}\right]_{3} , \nonumber   \\
 J_{U(1)_B}&=\left[-\frac{1}{2}\tilde{l}^{\dagger}\tilde{l}+\frac{1}{6}\tilde{q}^{\dagger}\tilde{q}+\frac{1}{3}\tilde{d}^{\dagger}\tilde{d}-\frac{2}{3}\tilde{u}^{\dagger}\tilde{u} +\tilde{e}^{\dagger}\tilde{e} \right]_{1,2}\,,  \\
 J^A_{SU(2)_A}&=\frac{1}{2}\left( H^{\dagger}_u\sigma^A H_u+ H^{\dagger}_d\sigma^A H_d\right)+\frac{1}{2}\left[\tilde{q}^{\dagger}\sigma^A \tilde{q} +\tilde{l}^{\dagger}\sigma^A \tilde{l}\right]_{3} \ ,  \nonumber \\
 J^A_{SU(2)_B}&=+\frac{1}{2}\left[\tilde{q}^{\dagger}\sigma^A \tilde{q} +\tilde{l}^{\dagger}\sigma^A \tilde{l}\right]_{1,2} .
 \end{align}
These results may be extended to a four Higgs doublet model or to a quiver model with three or more sites (see appendix \ref{app:Holography}), straightforwardly.
The sum in \eqref{superoperator} implies that the D-terms here described generate mass shifts to the Higgs doublets, to all charged squarks and sleptons (see appendix \ref{appendix:sfermionmatrix}) as well as additional quartic vertices. As a consequence, additional contributions to branching ratios should be considered in  precision studies with Higgs and sfermion decays.  An accurate detection of these effects may allow for the determination of the gauge structure and its matter charges identify the underlying model.

\section{The MSSM Higgs including vector type D-terms} \label{app:HiggsSfermions}  In this appendix we collect a number of relevant expressions for the Higgs sector with vector-like non-decoupling D-terms.  We will work in Feynman gauge, such that the $\xi$-terms are gauge-dependent contributions.
 For brevity we set \begin{align}
                     g_{12}^2&=\frac{3}{5}g_{1}^{2}(1+\Delta_1^{2}) + g_{2}^{2}(1+\Delta_2^{2})\,,\\
\hat{g}_{12}^2&=-\frac{3}{5}
g_{1}^{2}( 1 +\Delta_1^{2})  + g_{2}^{2}(1 +\Delta_2^{2})\,.
                    \end{align}
The non-decoupling D-terms of the vector type \refe{eq:nondecoupled}, appear in the
tadpole equations 
\begin{align} 
\frac{\partial V}{\partial H^0_{d}} &= \frac{1}{8} \Big(-8 v_u
\text{Re}[B_{\mu}] + g_{12}^2v_{d}^{3} + v_d [8 m_{H_d}^2  + 8
|\mu|^2  - g_{12}^2v_{u}^{2} ]\Big)\,, \\ 
\frac{\partial V}{\partial H^0_{u}} &= \frac{1}{8} \Big(-8 v_d
\text{Re}[B_{\mu}] + 8 v_u |\mu|^2  + v_u [8 m_{H_u}^2 
- g_{12}^2(- v_{u}^{2}  +
v_{d}^{2})]\Big)\,,
\end{align} 
as well as the Higgs mixing matrices. The mass matrix for the CP-even Higges, in the basis of the real components of $(H^0_d,\,H^0_u)$ is given by
\begin{equation} 
m^2_{h} = \!\!\left( 
\begin{array}{cc}
m_{h,11}&\!\!\!\! -\frac{1}{4} g_{12}^2 v_d v_u \! - \!\text{Re}[B_{\mu}]\\ 
\!\!\!-\frac{1}{4} g_{12}^2 v_d v_u \!\! - \text{Re}[B_{\mu}]  &m_{h,22}\end{array} 
\!\! \right) \,,
 \end{equation} 
where
\begin{align} 
m_{h,11} &= \frac{1}{8} \Big(8 m_{H_d}^2  + 8 |\mu|^2  + g_{12}^2\Big(3 v_{d}^{2}  -
v_{u}^{2} \Big)\Big)\,,\\ 
m_{h,22} &= \frac{1}{8} \Big(8 m_{H_u}^2  + 8 |\mu|^2  - g_{12}^2\Big(-3 v_{u}^{2}  +
v_{d}^{2}\Big)\Big)\,.
\end{align} 
For the pseudo-scalar Higgses, the mass matrix in the basis of the imaginary components of $(H^0_d,\,H^0_u)$ reads
\begin{equation} 
m^2_{A^0} = \left( 
\begin{array}{cc}
m_{A^0,11} &\text{Re}[B_{\mu}]\\ 
\text{Re}[B_{\mu}]&m_{A^0,22}\end{array} 
\right) +  \xi_{Z}m_Z^2 \,,
 \end{equation} 
where
\begin{align} 
m_{A^0,11} &= \frac{1}{8} \Big(8 m_{H_d}^2  + 8 |\mu|^2  + g_{12}^2\Big(- v_{u}^{2}  +
v_{d}^{2}\Big)\Big)\,,\\ 
m_{A^0,22} &= \frac{1}{8} \Big(8 m_{H_u}^2  + 8 |\mu|^2  - g_{12}^2\Big(- v_{u}^{2}  +
v_{d}^{2}\Big)\Big)\,.
\end{align} 
The mass matrix for the charged Higgses \( \left(H_d^-, H_u^{+,*}\right),
\left(H_d^{-,*}, H_u^+\right) \) reads
\begin{equation} 
m^2_{H^-} = \left( 
\begin{array}{cc}
m_{H^-,11} &\frac{1}{4} \Big(4 B_{\mu}^*  + \Big(g_{2}^{2} +
g_2^2\Delta_2^{2}\Big)v_d v_u \Big)\\ 
\frac{1}{4} \Big(4 B_{\mu}  + \Big(g_{2}^{2} + g_2^2\Delta_2^{2}\Big)v_d v_u
\Big) &m_{H^-,22}\end{array} 
\right) +  \xi_{W^-}m_{W^-}^2 \,,
 \end{equation} 
 with
\begin{align} 
m_{H^-,11} &= \frac{1}{8} \Big(8 m_{H_d}^2  + 8 |\mu|^2  + g_{12}^2v_{d}^{2}  + \hat{g}^2_{12} v_{u}^{2}
\Big)\,,\\ 
m_{H^-,22} &= \frac{1}{8} \Big(8 m_{H_u}^2  + 8 |\mu|^2  + g_{12}^2v_{u}^{2}  + \hat{g}^2_{12} v_{d}^{2}
\Big)\,.
\end{align} 
\section{The MSSM Higgs including chiral type D-terms}\label{app:HiggsSfermionschiral}
In this appendix we provide the relevant mass matrices and equations for the chiral type D-terms of \refe{eq:D-termsC}.
For the chiral type D-terms the tadpole equations are given by
\begin{align} 
\frac{\partial V}{\partial H^0_{d}} &= - v_u\text{R} \Big[B_{\mu}\Big] +v_d  \Big(m_{H_d}^2 + |\mu|^2\Big)+ v_d\frac{m_Z^2}{2}\cos(2\beta)+\frac{v_d}{8}  \sum_ik_ig_i^2\Omega_i\left(\frac{v_{d}^{2}}{\xi_i^2}+v_u^2\right)\,,\label{ChiralTadpoleAppendix1}\\ 
\frac{\partial V}{\partial  H^0_{u}} &=
 - v_d {\text{R}\Big[B_{\mu}\Big]+ v_u\Big( m_{H_u}^2+ |\mu|^2\Big)-v_u\frac{m_Z^2}{2}\cos(2\beta) }+\frac{v_u}{8}\sum_ik_ig_i^2\Omega_i\left(v_d^2+v_{u}^{2}\xi_i^2\right)\,.\label{ChiralTadpoleAppendix2}
\end{align} 
 where $k_i=(3/5,1)$. The mass matrix for the CP-even Higges, in the basis of the real components of $(H^0_d,\,H^0_u)$ is given by 
\begin{equation} 
m^2_{h} = \left( 
\begin{array}{cc}
m_{h,11} &  \frac{v_d v_u}{4} \sum_ik_ig_i^2(\Omega_i-1) -  \text{Re}[B_{\mu}] \\ 
 \frac{v_d v_u}{4} \sum_ik_ig_i^2(\Omega_i-1)  -  \text{Re}[B_{\mu}]  &m_{h,22}\end{array} 
\right)\,, 
 \end{equation} 
 with
\begin{align} 
m_{h,11} &=  |\mu|^2+ m_{H_d}^2+m_Z^2\left(\frac{\cos(2\beta)}{2}+\cos^2\beta\right)+\frac{1}{8}\sum_ik_ig_i^2\,\Omega_i\left(\frac{3v_{d}^{2}}{\xi_i^2}+v_u^2\right)\,,  \\ 
m_{h,22} &=|\mu|^2 + m_{H_u}^2-m_Z^2\left(\frac{\cos(2\beta)}{2}-\sin^2\beta\right)+\frac{1}{8}\sum_ik_ig_i^2\,\Omega_i\left(v_d^2+3\xi_i^2v_{u}^{2}\right) \,,
\end{align} 
that using the minimisation conditions \eqref{ChiralTadpoleAppendix1},\eqref{ChiralTadpoleAppendix2} become:
\begin{align} 
m_{h,11} &=  B_{\mu}\tan\beta+m_Z^2\cos^2\beta+\frac{v^2}{4}\cos^2\beta\sum_ik_ig_i^2\,\frac{\Omega_i}{\xi_i^2}  \\ 
m_{h,22} &=B_{\mu}\cot\beta+m_Z^2\sin^2\beta+\frac{v^2}{4}\sin^2\beta\sum_ik_ig_i^2\,\Omega_i\xi_i^2 \,.
\end{align} 
For the pseudo-scalar Higgses, the mass matrix in the basis of the imaginary components of $(H^0_d,\,H^0_u)$ reads
\begin{equation} 
m^2_{A^0} = \left( 
\begin{array}{cc}
m_{A^0,11} &{\text{R}\Big[B_{\mu}\Big]}\\ 
{\text{Re}\Big[B_{\mu}\Big]} &m_{A^0,22}\end{array} 
\right) +  \xi_{Z}m^2(Z) 
 \end{equation} 
 with
\begin{align} 
m_{A^0,11} &= |\mu|^2+ m_{H_d}^2  + \frac{m^2_Z}{2} \cos(2\beta)+\frac{1}{8}\sum_ik_ig_i^2\Omega_i\left(\frac{v_d^2}{\xi_i^2}+v_u^2\right)\,, \\ 
m_{A^0,22} &=  |\mu|^2+ m_{H_u}^2  + \frac{m^2_Z}{2} \cos(2\beta)+\frac{1}{8}\sum_ik_ig_i^2\Omega_i\left(v_d^2+v_u^2\xi_i^2\right) \,,
\end{align}
that using the minimisation conditions \eqref{ChiralTadpoleAppendix1},\eqref{ChiralTadpoleAppendix2} become:
\begin{align} 
m_{A^0,11} &=B_{\mu}\cot\beta\,,\\ 
m_{A^0,22} &=B_{\mu}\tan\beta\,.
\end{align}
The mass matrix for charged Higgses  \( \left(H_d^-, H_u^{+,*}\right), \left(H_d^{-,*}, H_u^+\right) \) is
 \begin{equation} 
m^2_{H^-} = \left( 
\begin{array}{cc}
m_{H^-,11} &\frac{1}{4} g_{2}^2(1-\Omega_2) v_d v_u + B_{\mu}^*\\ 
\frac{1}{4} g_{2}^{2}(1-\Omega_2) v_d v_u  + B_{\mu} &m_{H^-,22}\end{array} 
\right) +  \xi_{W^-}m^2(W^-) \,,
 \end{equation} 
 with
\begin{align} 
m_{H^-,11} &=  |\mu|^2+ m_{H_d}^2  +  \frac{v_u^2}{8} \Big[\frac{3}{5} g_1^2(\Omega_1-1)+g_2^2(1-\Omega_2) \Big]+\frac{v_d^2}{8}\sum_ik_ig_i^2\left(1+\frac{\Omega_i}{\xi_i^2}\right)\,,\\ 
m_{H^-,22}& = |\mu|^2 + \! m_{H_u}^2+\frac{v_d^2}{8}\left[ \frac{3}{5}g_1^2(\Omega_1 -1)+g_2^2(1-\Omega_2)  \right] +\frac{v_u^2}{8}\sum_ik_ig_i^2\left(\xi_i^2\Omega_i+1\right)\,,
\end{align} 
that using the minimisation conditions \eqref{ChiralTadpoleAppendix1},\eqref{ChiralTadpoleAppendix2} become:
\begin{align} 
m_{H^-,11} & =B_{\mu}\tan\beta+m_W^2\sin^2\beta(1-\Omega_2)\,,\\ 
m_{H^-,22}& =  B_{\mu}\cot\beta+m_W^2\cos^2\beta(1-\Omega_2)\,.
\end{align}

\section{Sfermion mixing matrices for the vector type with matter on site A}\label{appendix:sfermionmatrix}
The non-decoupling D-terms can have an effect also on the squark and slepton mixing matrices. For the simplest case where all MSSM-like matter including the Higgs is on site A, the mixing matrix  $M_{\tilde{f}}$ of a generic
sfermion $\tilde{f}$ for charged sleptons or squarks is given by
\begin{equation}
M^2_{\tilde{f}}=
\left( \begin{array}{cc}
m_{\tilde{f}_L}^2+m_{f}^2+\hat{M}_{Z}^2 (I^f_3-Q_f s_W^2) & m_f X^{\ast}_f  \\[.5em]
m_f X_f  & m_{\tilde{f}_R}^2+m_f^2+\hat{M}_{Z}^2\,Q_f s_W^2
\end{array} \right),
\label{eq:sfermion}
\end{equation}
denoting $s_w=\sin\theta_W$ where $\theta_W$ is the Weinberg weak mixing angle, 
and the useful abbreviation $\hat{M}_{Z}^2\equiv (m_Z^2+m_{\Delta}^2)\cos{2\beta}$, where $m_{\Delta}^2=\frac{1}{2}(\frac{3}{5}g_1^2\Delta_1+g_2^2\Delta_2)v^2$.
The off-diagonal element $X_f$ is defined in terms of the soft SUSY-breaking trilinear
coupling $A_f$ via
\begin{equation}
 X_f=A_f-\mu^{\ast} \left\{\cot\beta,\tan\beta\right\},
\end{equation}
where $\cot\beta$ applies for the up-type quarks, $f=u,c,t$, and
$\tan\beta$ applies for the down-type fermions, $f=d,s,b,e,\mu,\tau$. 
Note that $m_f$, $Q_f$ and $I_3^f$ are the mass,
charge and isospin projection 
of the fermion $f$, respectively.
Once diagonalised this matrix leads to the light and heavy sfermion
masses $m_{\tilde{f}_1}$ and $m_{\tilde{f}_2}$. In particular the stop masses are given by
\begin{align}
 m_{\tilde{t}_{1,\,2}}^2=&m_t^2+\frac{1}{2}\left[M_{\tilde{Q_3}}^2+M_{u_3}^2+\frac{1}{2}\hat{M}_Z^2\cos2\beta\right.\nonumber\\
 &\left.\mp\sqrt{\left[M_{Q_3}^2-M_{u_3}^2+\hat{M}_Z^2\cos2\beta\left(\frac{1}{2}-\frac{4}{3}\sin^2\theta_W\right)\right]^2+4m_t^2X_t^2}\right]\,.\label{SfermionMassesMatrix}
\end{align}
To obtain the MSSM mass expression one has just to set $m_{\Delta}=0$.
For a light stop scenario, this may have an appreciable effect and similarly, for the stau which may be the NLSP (for a Goldstino LSP scenario such as GMSB).  In the case that the $g_A>g_B$, even for the case of split families the above mixing matrix will still apply to the third generation scalars on Site A. For the sneutrinos the mass matrix is given by
\begin{equation}
 m_{\tilde{\nu}}^2= M_L^2 + \frac{1}{2}(m_Z^2+m_{\Delta}^2)\cos(2\beta).
\end{equation}


%

\section{Perturbative unification and the size of the D-terms}\label{sec:unification}
\begin{figure}[ht!]
\begin{center}
\includegraphics[scale=0.5]{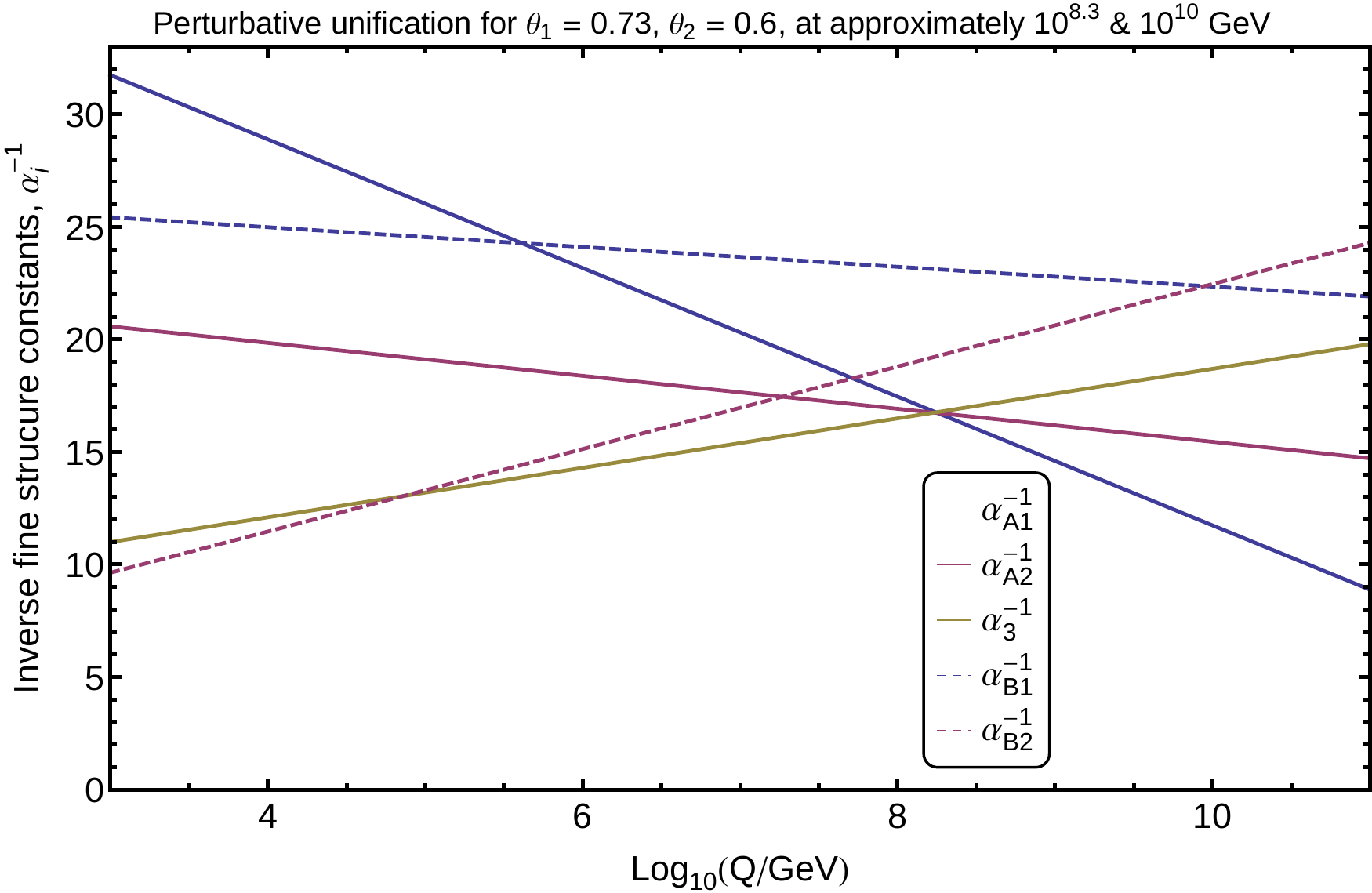}
\caption{Perturbative unification of the $G_A\times SU(3)_c$ and $G_B$ sites separately, allowing for the maximal value of the ratios $R_i$.  These also give a prediction of the values of the mixing angles, $\theta_i$'s in \refe{thetas}. } 
\label{fig:RGEs}
\end{center}
\end{figure}
To maximise the effect of the vector-like D-terms such as \refe{eq:nondecoupled}, one requires that the ratio of gauge couplings
\be
R_i=g_{Ai}/g_{Bi}\,,
\ee
is as large as possible, however making certain gauge couplings large at low energies and including additional matter fields will certainly effect perturbativity of the gauge couplings at higher energies.  In addition, whilst these models do not (yet) have full GUT multiplets of matter, particularly for the linking fields, but also for the MSSM matter content, we can still explore the possibility of unification in these models as usual.  For definiteness we take the model outlined in table \ref{Table:matterfieldsVeCModelA}.  

The beta functions at one loop are given by
\be
\beta_{g_a}=\frac{d}{dt}g_a=\frac{b_a}{16\pi^2}g^3_a  \quad \text{with} \quad   b_a=(2,\frac{39}{5},-5,\frac{6}{5},-3)\,,
\ee
The restriction that $\alpha_i(M_{GUT})<1$ and that
\be 
\alpha_{g_{1A}}(M_{GUT})=\alpha_{g_{2A}}(M_{GUT})=\alpha_{g_c}(M_{GUT})\,,\ee
with 
\be 
\alpha_{g_{1B}}(M_{GUT})=\alpha_{g_{2B}}(M_{GUT})\,,
\ee
restricts the parameter space significantly. The results of perturbative unification for the largest values of R's are plotted in figure \ref{fig:RGEs}. We find $R_1\sim 0.6$ and $R_2\sim 0.86$, such that even allowing for $\frac{m_L^2}{m_{v}^2+m_{L}^2}\sim 1$ this gives
\be
\Delta=\left(\frac{g^2_{A}}{g^2_{B}}\right)\frac{m_L^2}{m_{v}^2+m_{L}^2} < R  \quad \text{which implies that}  \quad (\Delta^{Max}_1,\Delta^{Max}_2)=  (0.6,0.86)\,,
\ee
 respectively.  Larger $\Delta$'s are also possible if $SU(3)$ is quivered to $SU(3)_A\times SU(3)_B$ as then $\alpha^{-1}_{3A}$ can be made weaker allowing for unification at a later scale and therefore larger $R$'s.
Of course abandoning perturbativity altogether will allow for a larger D-term enhancement too. 

\begin{figure}[ht!]
\begin{center}
\includegraphics[scale=0.5]{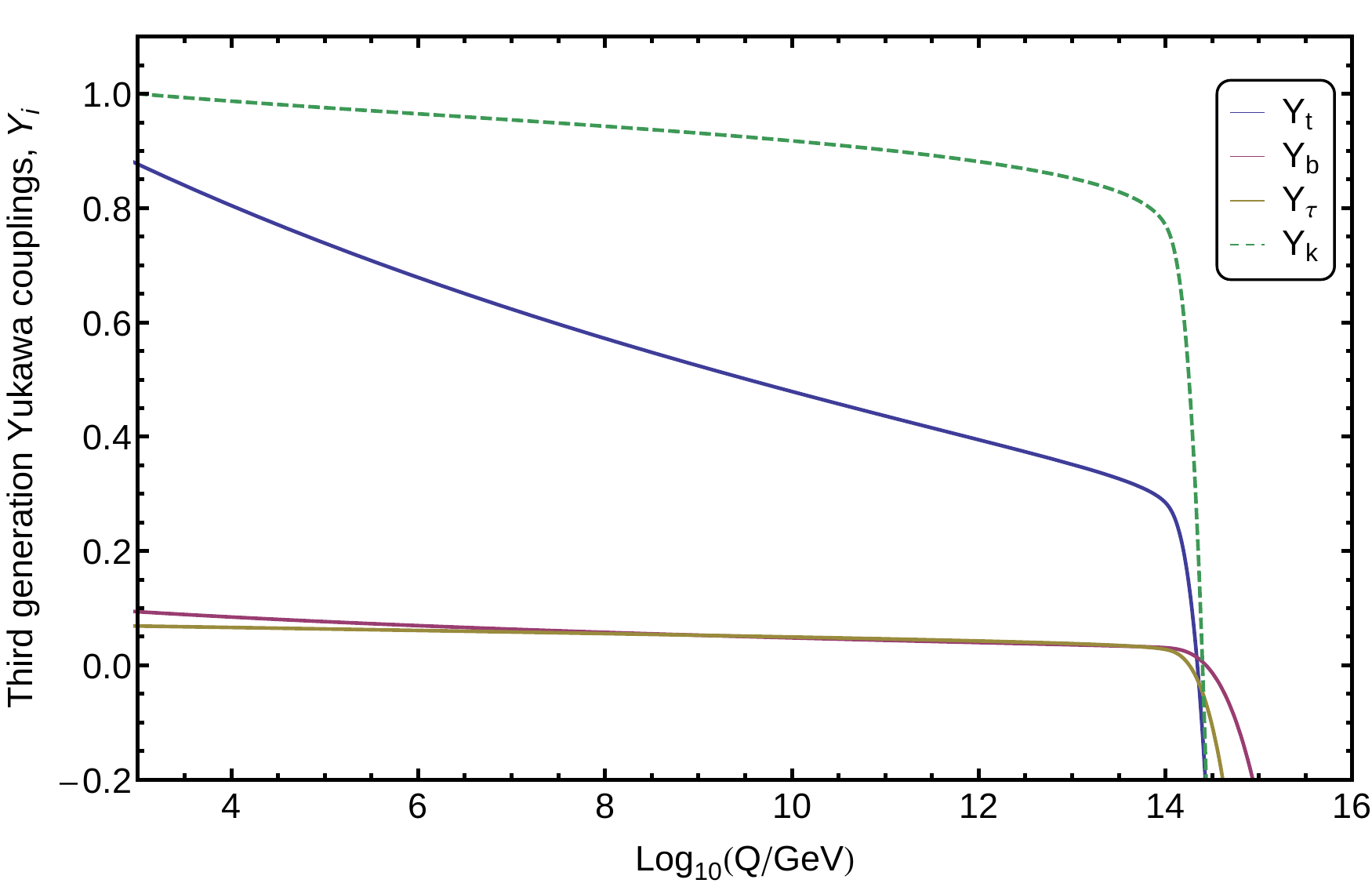}
\caption{Renormalisation group evolution for the Yukawa couplings.  } 
\label{fig:Yuks}
\end{center}
\end{figure}

The dynamics of the Yukawa couplings are also of interest The superpotential is given by
\be
W=W_{MSSM}+ \frac{Y_k}{2} K (L\tilde{L}-V^2)\,,
\ee
in which $Y_k$ is taken to be real. The Yukawa coupling one-loop beta functions (for the third generation only) are given by
\be
\beta^{(1)}_{y_t}\equiv\frac{d}{dt}y_{t}\simeq \frac{y_t}{16\pi^2}\left[ 3y_t^* y_t 
+y^*_by_b-\frac{16}{3}g_3^2 -3g^2_{A2}-\frac{13}{15}g^2_{A1}\right],
\ee
\be
\beta^{(1)}_{y_b}\equiv\frac{d}{dt}y_{b}\simeq \frac{y_b}{16\pi^2}\left[ y_t^* y_t +y_{\tau}^* y_{\tau}
+3y^*_by_b-\frac{16}{3}g_3^2 -3g^2_{A2}-\frac{7}{15}g^2_{A1}\right],
\ee
\be
\beta^{(1)}_{y_{\tau}}\equiv\frac{d}{dt}y_{\tau}\simeq \frac{y_{\tau}}{16\pi^2}\left[ 4y_{\tau}^* y_{\tau}+ y_t^* y_t
-\frac{16}{3}g_3^2 -3g^2_{A2}-\frac{7}{15}g^2_{A1}\right],
\ee
and
\be
\beta^{(1)}_{y_{k}}\equiv\frac{d}{dt}y_{k}\simeq \frac{y_{k}}{16\pi^2}\left[ \frac{15}{10}y_{k}^* y_{k} -10g^2_{B2}-10g^2_{A2}-6g^2_{A1}-6g^2_{B1}\right].
\ee
For the same choice of parameters as before, the results are presented in figure \ref{fig:Yuks}, where one can see that as $\alpha_{A1}$ hits a Landau pole at around $10^{14}$ GeV, which is after both GUT scales,  the Yukawa couplings become very small and run to opposite signed values.  This is reminiscent of five dimensional extensions of the MSSM \cite{Dienes:1998vh,Abdalgabar:2014bfa}, where power law running is used to argue for an explanation of the $O(1)$ top Yukawa from an initially small coupling in the UV.

\section{Other quiver alternatives}\label{sec:otherquivers}
In this appendix we introduce a number of extensions of the quiver models outlined in this paper.  First we mention a four Higgs doublet model that may lead the 2HDM with chiral Higgs doublets described in sec. \ref{subsec:ChiralCase}. Then we discuss a three site quiver that ``deconstructs" a holographic extra dimension and may also help to explain the flavour hierarchies of the SM and give a natural SUSY hierarchy of light 3rd generation squarks \cite{Craig:2012hc,Bharucha:2013ela}.

\subsection{Four Higgs doublet model (4HDM)}\label{subsec:4HDM}

\begin{figure}[t!]
\begin{center}
\includegraphics[scale=0.5]{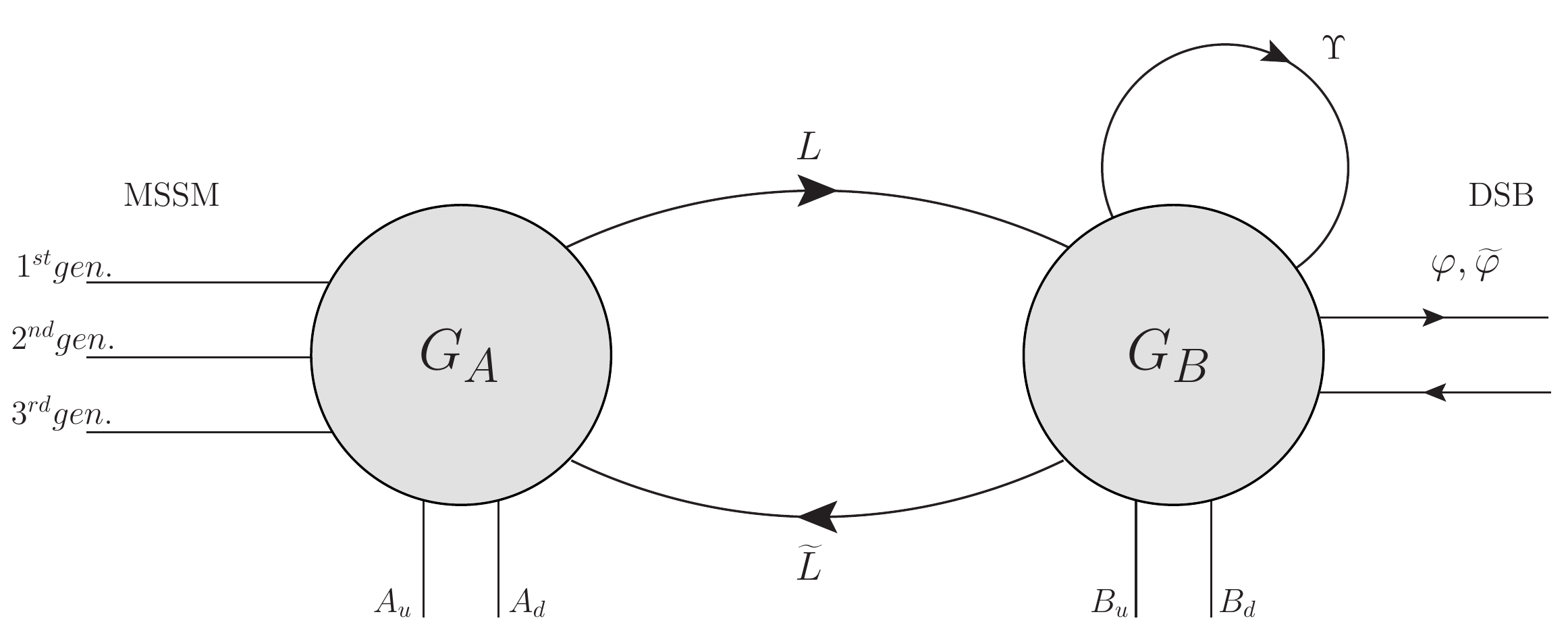}
\caption{An example of UV completion of the model in figure \ref{fig:ModelC} of the chiral-Higgs like non-decoupling D-term.  This is a four Higgs doublet model in which the Higgses $A_u$ and $B_d$ combine to make the MSSM Higgses $H_u,H_d$ with additional chiral-Higgs non-decoupling D-terms in their scalar potential, while the remaining two states are integrated out at a suitable scale to give the D-terms of \refe{eq:D-termsC}.  It is also possible that all four Higgses have masses below the linking scale, in which case one obtains a four Higgs doublet variant of the MSSM with D terms as in \refe{eq:D-termsD}. The $\Upsilon$ adjoint field and the messenger fields $\varphi, \tilde{\varphi}$ are coupled to site B.} 
\label{fig:UVofModelC}
\end{center}
\end{figure}

An alternative possibility to the MSSM is a model with four Higgs doublets below the scale of the linking field vevs, see figure \ref{fig:UVofModelC}. Labelling the $ A$ and $B$ site Higgses $(A_u,A_d,B_u,B_d)$, in this case the non-decoupling D-terms would take the form
\begin{alignat}
\delta \mathcal{L}=&-\frac{3}{5}\frac{g_1^2\Omega_1}{8}  \left( \xi_1 (A^{\dagger}_u
A_u-A_d^{\dagger}A_d)- \frac{1}{\xi_1} (B^{\dagger}_u
B_u-B_d^{\dagger}B_d)\right)^2\nonumber \\
&
-\frac{g_2^2\Omega_2 }{8}\sum_a
\left( \xi_2(A^{\dagger}_u\sigma^a A_u+A^{\dagger}_d\sigma^a A_d )-\frac{1}{\xi_2}(B^{\dagger}_u\sigma^a B_u+B^{\dagger}_d\sigma^a B_d ) \right)^2+\dots\,\,.\label{eq:D-termsD}
\end{alignat}
where the ellipsis denote other squark and slepton contributions to the D-term potential.

\subsection{Deconstructed Holography} \label{app:Holography}
\begin{figure}[t!]
\begin{center}
\includegraphics[scale=0.6]{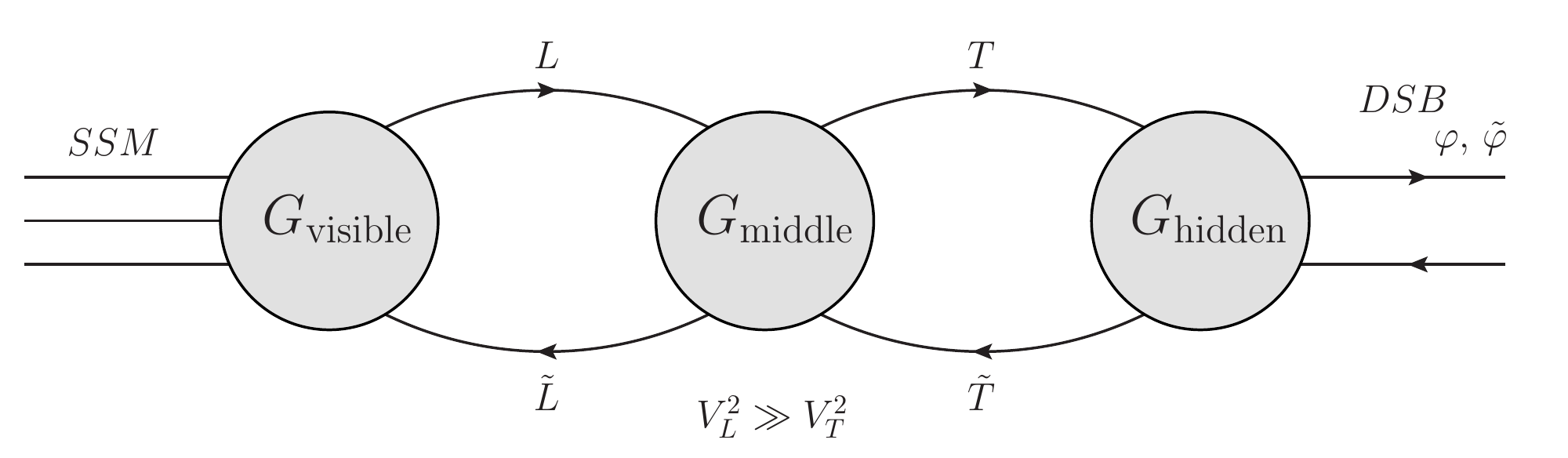}
\caption{A 3 site model of a deconstructed holographic setup. The vevs of the linking fields encode a discretized metric. The non-decoupling D-term of the 3-site case is more exotic to derive but it may be approximated by the 2-site case between each pair of gauge sites, by integrating out $(L,\tilde{L})$ and $(T,\tilde{T})$ separately.} 
\label{fig:Modelwarped}
\end{center}
\end{figure}

A number of models of supersymmetry breaking involve strong coupling or holography and this may be usefully approximated by holographic deconstruction \cite{Randall:2002qr,Falkowski:2002cm}, its most elementary example is given by a 3-site quiver model. Such models may also exhibit non-decoupling D-terms, as well as a possible explanation of flavour hierarchies and a squark soft mass hierarchy, which motivates giving them a brief mention.  The key to realising such a scenario is to identify the metric
\begin{equation}
 ds^2= e^{-2 k y }\eta_{\mu\nu}dx^{\mu}dx^{\nu}- dy^2\,.
\end{equation}
The relative vevs of the linking fields related as
\be
\frac{v_T}{v_L}=\frac{e^{-2\sigma_i}a^2_L}{a^2_T}\,,
\ee
where the vevs of the linking fields are labelled $v_T,v_L$ as in figure \ref{fig:Modelwarped} and the lattice spacings are $a_T,a_L$. A useful identification is to take the lattice spacing and couplings equal $a_i=a$ and $g_i(v_i)=g$, such that the warping is entirely encoded in the linking field vevs. Naturally one would find that 
\be
m_{v_L}\gg m_{v_T} \,,
\ee
in which case by inspecting \refe{eq:nondecoupled} the prevalent effect of the non-decoupling D-term would be on matter close to the IR part of the quiver where the effect of the supersymmetry breaking would be maximal. One may therefore approximate the 2-site non-decoupling D-term as arising from integrating out the linking fields between $G_{middle}$ and $G_{hidden}$ as in figure \ref{fig:Modelwarped} (for a scenario in which the supersymmetry breaking is predominately gauge mediated from fields charged under $G_{hidden}$).  

Even in the case of $v_T\sim v_L$ (the flat case) the three site model has a number of interesting features. A possible model of soft mass hierarchies would be to locate each of the three generations of MSSM-like matter on the three separate gauge sites, respectively.  Superfields further away from the source of supersymmetry breaking would then have smaller soft masses, compared to those located closest to the supersymmetry breaking effects \cite{McGarrie:2010qr}.  In addition one may attempt to explain the flavour hierarchies, if for example the $H_u,H_d$ are on the same site as the 3rd generation superfields, this might lead to an O(1) Yukawa coupling, but the subleading Yukawa entries would only be generated through nonrenormalisable operators from integrating out both sets of linking fields. 


\bibliographystyle{JHEP}
\bibliography{Dterms}

\end{document}